%% file: draft.tex
\documentclass[a4paper,11pt]{article}
\pdfoutput=1 % if your are submitting a pdflatex (i.e. if you have
\usepackage{jcappub}
\usepackage{graphicx}
\usepackage{color}
\usepackage{makecell}
\usepackage{afterpage}
\usepackage[dvipsnames]{xcolor}
\usepackage{multirow}
\usepackage{cleveref}
\usepackage[normalem]{ulem}		% Automatic eq. fig., etc.
\usepackage{xspace}
\usepackage{subcaption}
\usepackage{booktabs}
\usepackage{centernot}
\usepackage{mathtools}
\usepackage{stmaryrd}
\usepackage[utf8]{inputenc}
\usepackage{csquotes} % For \enquote{}
\usepackage{placeins}
\usepackage{titlesec}
\usepackage{enumitem}

\newcommand{\class}{{\sc class}\xspace}
\newcommand{\lcdm}{\ensuremath{\Lambda\mathrm{CDM}}\xspace}
\newcommand{\lya}{Lyman-$\alpha$}
\newcommand{\rr}{\mathrm}

\newcommand{\comm}[1]{}

\title{One likelihood to bind them all: Lyman-$\alpha$ constraints on non-standard dark matter}

\author[1]{Deanna C. Hooper,}
\author[2]{Nils Sch\"oneberg,}
\author[3,4]{Riccardo Murgia,}
\author[5,6]{Maria Archidiacono,}
\author[7]{Julien Lesgourgues,}
\author[8,9,10,11]{and Matteo Viel}

\affiliation[1]{Department of Physics and Helsinki Institute of Physics, PL 64, FI-00014 University of Helsinki, Finland}
\affiliation[2]{Institut de Ci\`encies del Cosmos, Universitat de Barcelona, Mart\'{\i} i Franqu\`es 1, Barcelona 08028, Spain}
\affiliation[3]{Gran Sasso Science Institute (GSSI), Viale F. Crispi 7, L’Aquila (AQ), I-67100, Italy}
\affiliation[4]{INFN - Laboratori Nazionali del Gran Sasso (LNGS), L’Aquila (AQ),  I-67100, Italy}
\affiliation[5]{Dipartimento di Fisica ``Aldo Pontremoli", Universit\`a degli Studi di Milano, via G. Celoria 16, 20133 Milano, Italy}
\affiliation[6]{INFN - Sezione di Milano, Via G. Celoria 16, 20133 Milano, Italy}
\affiliation[7]{Institute for Theoretical Particle Physics and Cosmology (TTK), RWTH Aachen University, D-52056 Aachen, Germany.}
\affiliation[8]{SISSA, Via Bonomea 265, 34136 Trieste, Italy}
 \affiliation[9]{INFN, Sez. di Trieste, Via Valerio 2, 34127 Trieste, Italy}
 \affiliation[10] {IFPU, Institute for Fundamental Physics of the Universe, via Beirut 2, 34151 Trieste, Italy}
\affiliation[11]{INAF,  Osservatorio Astronomico di Trieste, via Tiepolo 11, I-34131 Trieste, Italy}

\emailAdd{deanna.hooper@helsinki.fi}
\emailAdd{nils.science@gmail.com}
\emailAdd{riccardo.murgia@gssi.it}

\abstract{ 
Recent cosmological tensions have rekindled the search for models beyond $\Lambda$CDM that cause a suppression of the matter power spectrum. Due to the small scales accessible to  Lyman-$\alpha$ data they are an excellent additional tool to probe such models. In this work we extend a recently-developed approach for using Lyman-$\alpha$ data to constrain the power spectrum suppression caused by almost any mixture of cold and non-standard dark matter. We highlight the steps involved in the development of a corresponding likelihood that will be publicly released upon publication of this work. We study three examples of models suppressing the power spectrum, namely feebly interacting dark matter, dark matter interacting with baryons, and mixed cold+warm dark matter. The latter two can be well constrained from Lyman-$\alpha$ data, and we derive novel conclusions on the cosmologically allowed parameter spaces, including finding a mild preference for non-zero interactions between dark matter and baryons. The consistency of the constraints obtained on these models highlight the robustness and flexibility of the likelihood developed here.
}

\begin{document}

\hfill{\small HIP-2022-17/TH}\\
\vspace{-0.5 cm}
\hfill{\small TTK-22-21}

\maketitle

\input{Introduction}
\input{Lya}
\input{Models}
\input{Results}
\input{Conclusions}

\section*{Acknowledgements}
The authors thank M. Lucca for insightful discussions and useful feedback on the manuscript. DH is supported by the Academy of Finland grant no. 328958. NS acknowledges support from the DFG grant LE 3742/4-1. NS also acknowledges the support of the following Maria de Maetzu fellowship grant: Esta publicaci\'on es parte de la ayuda CEX2019-000918-M, financiado por MCIN/AEI/10.13039/501100011033. MV is supported by the INFN PD51 INDARK grant. Simulations were performed with computing resources granted by JARA-HPC from RWTH Aachen University under project jara0184, on the Ulysses SISSA/ICTP supercomputer, and further computational resources have been provided by the CÉCI, funded by the F.R.S.-FNRS under Grant No. 2.5020.11 and by the Walloon Region.
\input{Appendix}

\FloatBarrier
\bibliography{biblio}{}
\bibliographystyle{JHEP}

\end{document}

%% file: Introduction.tex
\section{Introduction}
\label{sec:intro}

Cold, collisionless dark matter (CDM) is an essential ingredient of the standard cosmological model, $\Lambda$CDM. This type of dark matter (DM) is characterised by its relatively high mass and lack of significant non-gravitational interactions with Standard Model particles. The existence of CDM is supported by observations ranging from the Cosmic Microwave Background (CMB) radiation~\cite{Akrami:2018vks}, to Baryon Acoustic Oscillations (BAO)~\cite{Beutler:2011hx,Ross:2014qpa,Alam:2016hwk}, to the distribution of matter on large scales~\cite{Clowe:2006eq,Heymans:2013fya,Abbott:2020knk,Joudaki:2019pmv}. Despite this, the most well motivated DM candidate, known as Weakly Interacting Massive Particles (WIMPs) -- where the DM is coupled weakly to the Standard Model and produced via a freeze-out mechanism -- has remained elusive in direct, indirect, and collider searches.

Furthermore, despite the remarkable success of $\Lambda$CDM in explaining many cosmological observables, some issues remain. On the one hand, there is a persistent tension between the $\Lambda$CDM-predicted value~\cite{Akrami:2018vks,Beutler:2011hx,Ross:2014qpa,Alam:2016hwk} of the expansion rate of the universe, $H_0$\,, and the its direct measurements via standard candles~\cite{Knox:2019rjx,Riess:2019cxk,Verde:2019ivm,DiValentino:2021izs,Schoneberg:2021qvd, Riess:2021jrx}. 
A similar, albeit less significant, tension also appears between the expected value~\cite{Akrami:2018vks} of the clustering of matter on scales of $8\,\text{Mpc}/h$, $\sigma_8$\,, and the measurements obtained with weak lensing experiments~\cite{Heymans:2013fya,Asgari:2019fkq,Abbott:2020knk,Joudaki:2019pmv,2020arXiv200715632H}. Indeed, a tension between the inferred  $\sigma_8$ values of galaxy cluster number counts and the values inferred from a Lyman-$\alpha$ forest analysis based on 1D flux power of the same data set considered in this paper \cite{esposito22} was recently pointed out.
Furthermore, there are discrepancies in the distribution of matter on small scales when comparing our simulations of CDM to local observations~\cite{Flores:1994gz,Moore:1994yx,Klypin:1999uc,deBlok:2009sp,BoylanKolchin:2011de,BoylanKolchin:2011dk, Oman:2015xda,Kamada:2016euw,Tulin:2017ara,Salucci:2018hqu}. Moreover, in 2018 the EDGES collaboration reported a substantially colder 21cm spin temperature than expected~\cite{Bowman:2018yin} (see however Ref.~\cite{Singh:2021mxo}). Combined, these discrepancies have fuelled interest in DM models beyond the standard CDM, which we will refer to as non-standard dark matter (NSDM).

The most well-known NSDM model is warm dark matter (WDM), where thermalised DM freezes out relativistically in the early universe, substantially impacting late-time structure formation. Such models in which all or a fraction of the DM is warm (known as mixed WDM+CDM) have been studied extensively in the literature~\cite{Viel:2005qj, Boyarsky:2008xj, Viel:2013fqw, Baur:2017stq, Palanque-Delabrouille:2019iyz, Garzilli:2019qki,Enzi:2020ieg}.
Another broad class of NSDM, which has received a lot of attention in the literature, is the possibility of DM having non-gravitational interactions either with Standard Model particles, such as 
photons~\cite{Boehm:2004th,Weiner:2012cb,Wilkinson:2013kia,Boehm:2014vja,Ali-Haimoud:2015pwa,Diacoumis:2018ezi,Escudero:2018thh,Kumar:2018yhh,Stadler:2018jin,Ali-Haimoud:2021lka},
baryons~\cite{Chen:2002yh,Boehm:2004th,Sigurdson:2004zp,Melchiorri:2007sq,ArkaniHamed:2008qn,Dvorkin:2013cea,Ali-Haimoud:2015pwa,Munoz:2015bca,Kadota:2016tqq,Munoz:2017qpy,Gluscevic:2017ywp,Boddy:2018kfv,Ali-Haimoud:2018dvo,Barkana:2018lgd,Boddy:2018wzy,Bringmann:2018cvk,Emken:2018run,Slatyer:2018aqg,Xu:2018efh,Maamari:2020aqz,Ali-Haimoud:2021lka,Buen-Abad:2021mvc,Nguyen:2021cnb, Rogers:2021byl}, 
neutrinos~\cite{Bringmann:2013vra,Audren:2014lsa,Cherry:2014xra,Wilkinson:2014ksa,Horiuchi:2015qri,Ghosh:2017jdy,Diacoumis:2018ezi,Campo:2018dfh,DiValentino2018a,Olivares-DelCampo2018,Pandey:2018wvh,Choi:2019ixb, Stadler:2019dii,Green:2021gdc,Mosbech:2020ahp,Hooper:2021rjc}, 
or multiple of these simultaneously~\cite{Becker:2020hzj}; 
or within the dark sector~\cite{Murgia:2016ccp,DiValentino:2019ffd,Lucca:2020zjb,Lucca:2021dxo,Amiri:2021kpp,Gariazzo:2021qtg}, with a relativistic component such as
dark radiation~\cite{Archidiacono:2011gq, Diamanti:2012tg, Cyr-Racine:2013fsa,Chu:2014lja,Rossi:2014nea,Buen-Abad:2015ova,Lesgourgues:2015wza,Cyr-Racine:2015ihg,Schewtschenko:2015rno,Krall:2017xcw,Archidiacono:2017slj,Buen-Abad:2017gxg,Archidiacono:2019wdp}, 
or even DM self-interactions~\cite{Spergel:1999mh,ArkaniHamed:2008qn,Feng:2009mn,Feng:2009hw,Buckley:2009in,Tulin:2017ara}. 
NSDM models offer many possibilities to address the aforementioned tensions, although to date no model has been able to satisfactorily address all of the open issues within $\Lambda$CDM. A common element of many of these models is their impact on the matter power spectrum, where the warmness or added interactions induce a suppression of power on small scales. As such, small scale structure data is crucial to constrain these models. 

In this paper we will use data from the flux power spectra extracted from the so-called Lyman-$\alpha$ forest, which arises from many redshifted absorption lines in the spectra of distant quasars due to the intervening filaments of neutral hydrogen in the Inter-Galactic Medium (IGM). 
As these hydrogen filaments preferably cluster in the potential wells of the DM, they are an excellent tracer of the latter~\cite{Ikeuchi1986, 10.1093/mnras/218.1.25P,McQuinn:2015icp}. Additionally, due to the high spectroscopic resolution of the  
instruments, Lyman-$\alpha$ data probe structure formation on comparatively small scales. However, the flux observations relate to the underlying matter distribution in a highly non-linear way, and thus an analysis usually requires dedicated computationally demanding hydrodynamical simulations in order to establish the corresponding relation and correct for the influence of the thermal state of the filaments~\cite{bolton17}. 
A novel approach to circumvent the need for suites of simulations was proposed in Refs.~\cite{Murgia:2017lwo, Murgia:2018now}, and a subsequent {\sc MontePython}~\cite{Audren:2012vy,Brinckmann:2018cvx} likelihood was developed in Ref.~\cite{Archidiacono:2019wdp}. Here we extend this approach to encompass a wider variety of DM models, allowing for suppressions in the matter power spectrum which induce a plateau at large $k$ instead of dropping to zero. This approach is especially suited for models in which the NSDM represents only a fraction of the total DM.

\enlargethispage*{2\baselineskip}
Here we use this newly-developed Lyman-$\alpha$ likelihood, together with the version of {\sc class}~\cite{Blas:2011rf} developed in Ref.~\cite{Becker:2020hzj} (which incorporates interacting DM), to constrain four different scenarios: mixed \mbox{WDM+CDM}; two models of interacting DM with baryons (specifically with protons) via a small but non-negligible velocity-dependent coupling; and the case of feebly interacting DM with dark radiation, in which we have an additional relativistic dark component with very weak interactions with the DM.

This paper is organised as follows. We introduce our new \lya~likelihood and discuss its scope and limitations in Sec.~\ref{sec:lya}. We then briefly review the different NSDM models considered here in Sec.~\ref{sec:models}.
In Sec.~\ref{sec:results} we discuss the results of our MCMC analyses, presenting new constraints on these models, before concluding in Sec.~\ref{sec:conc}.

%% file: Lya.tex
\section{Lyman-\texorpdfstring{$\alpha$}{alpha} constraints on exotic models}\label{sec:lya}

The Lyman-$\alpha$ forest is a collection of absorption lines in the spectra of distant quasars. The light coming from these quasars travels through intermediate regions containing neutral hydrogen, where it can be locally absorbed through the Lyman-$\alpha$ transition. As such, the positions of the absorption lines are an excellent tracer of the intervening neutral hydrogen in the universe. The Lyman-$\alpha$ forest is one of the most precise high-redshift and small-scale measurements currently available to cosmology, see e.g. Refs.~\cite{Weinberg:2003eg,McQuinn:2015icp} for a review.

\subsection{Lyman-$\alpha$ data from MIKE/HIRES}

The HIRES (KECK) and MIKE (Magellan) instruments are precise echelle spectrographs with resolution factors\footnote{Defined here as $1/\delta \nu = \frac{\lambda}{c \delta \lambda}$ for some observed wavelength $\lambda$ and corresponding uncertainty $\delta \lambda$.} of 0.074\,s/km and 0.14\,s/km, respectively \cite{Viel:2013apy} (to be compared to eBOSS with $\sim$0.014\,s/km \cite{Chabanier:2018rga,duMasdesBourboux:2020pck}). In particular, we consider four\footnote{We do not use the highest redshift bin for MIKE data, for which the error bars are very large~\cite{Viel:2013apy}.} redshift bins $z = \{4.2, 4.6, 5.0, 5.4\}$ and 10 wavenumber bins in the interval 0.005\,s/km - 0.08\,s/km. As in previous analyses \cite{Viel:2013apy,Irsic:2017ixq,Irsic:2017yje,Kobayashi:2017jcf,Nori:2018pka,Murgia:2019duy,Archidiacono:2019wdp,Miller:2019pss,Enzi:2020ieg}, we impose the cut of $k > 0.005$ s/km to suppress possible systematic effects from the continuum fitting procedure.

While far fewer quasars are observed relative to low-resolution surveys such as BOSS/eBOSS, it is important to realise that the high-resolution data are derived both at higher wavenumbers and higher redshift. This leads to overall stronger constraints on the models under consideration, which usually exhibit a suppression of the power at smaller scales. In this sense, the two kinds of surveys are highly complementary.

\subsection{Exotic models and power spectrum suppression}\label{ssec:suppression}

Our aim is to use the excellent precision of high-redshift Lyman-$\alpha$ data in order to constrain the parameter space of several NSDM relics. Therefore, we are interested in quantifying the suppression of the Lyman-$\alpha$ flux power spectrum caused by such a relic. 
To this end, and as has been shown extensively in the literature \cite{Murgia:2017cvj,Murgia:2017lwo,Murgia:2018now,Murgia:2019duy}, it can be very useful to describe the shape of suppression of the linear power spectrum of density perturbations with a generic formula.
Below we attempt to give a general reasoning for what kind of suppression shapes are expected.

\enlargethispage*{1\baselineskip}
A first step to discuss a suppression is defining with respect to what the power spectrum is suppressed, for which we take here a `reference' $\Lambda$CDM model, which in our case is chosen to have the same asymptotic limit for small $k \ll k_\mathrm{eq}$\,.\footnote{In practice, we use a different convention for each of the investigated models. For a mixed cold+warm model (\cref{sec:mixed}), we convert the NSDM content into CDM, whereas for the two models of interaction (\cref{sec:dmb,sec:fidm}) we simply set the corresponding interaction strength to zero. Otherwise, we keep all other parameters (such as for example $\omega_b\,, H_0$) the same. This effectively gives models with the same large-scale behaviour (and thus the same power spectrum for $k \ll k_\mathrm{eq}$).}

Theoretically the suppression in the matter power spectrum caused by the presence of some NSDM relic could depend on the redshift under consideration. However, for many models of DM interactions (such as those we investigate within the context of this work), the suppression relative to a reference $\Lambda$CDM model stays constant with respect to redshift. This is because the scattering (or thermal free-streaming) usually becomes irrelevant even before recombination, or otherwise has very little redshift-dependence. Due to the almost scale-independent growth after the initial imprint of such a suppression, the shape therefore stops varying at low redshifts relevant for our analysis. 

\begin{figure}[t]
	\centering
	\includegraphics[width=0.8\textwidth]{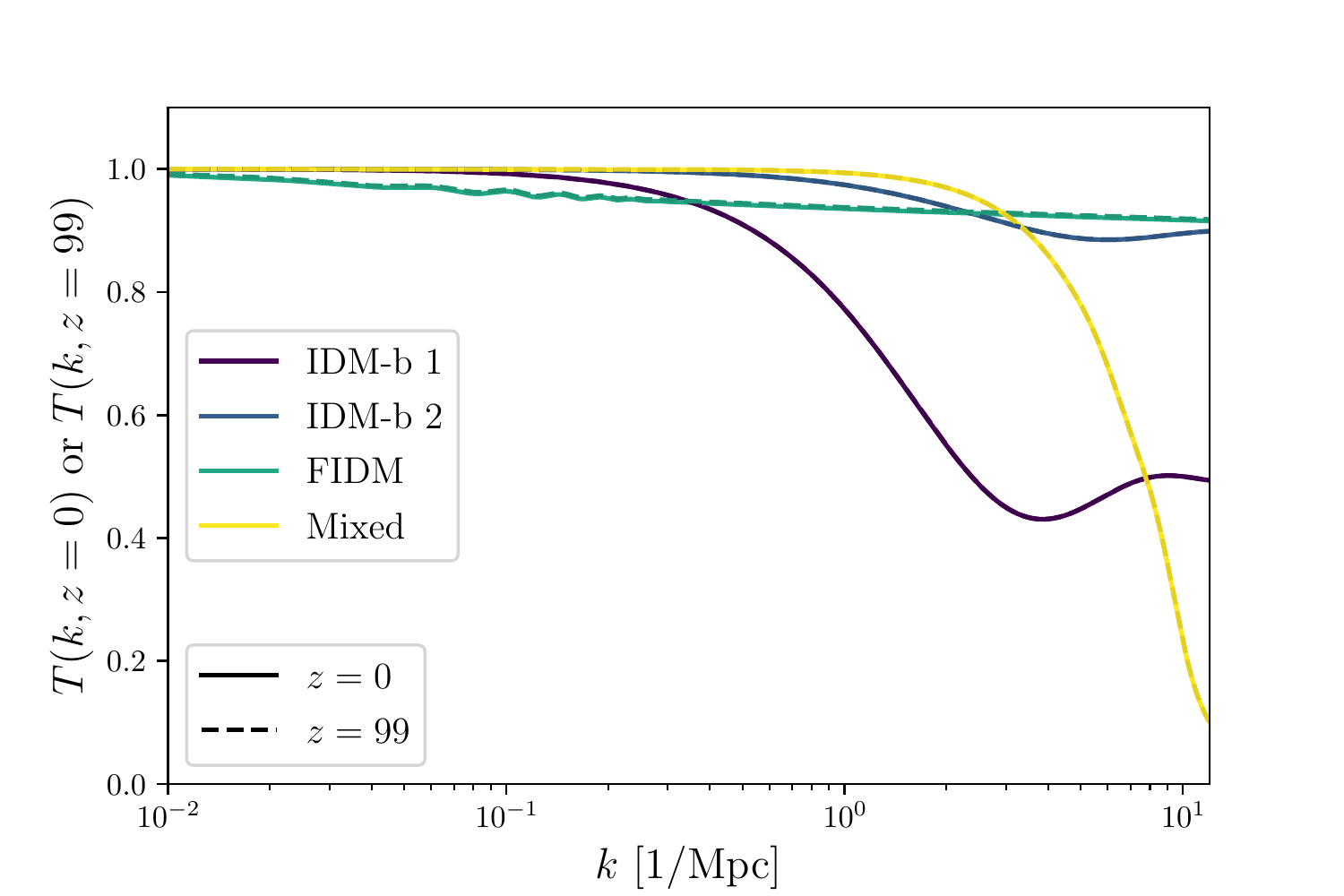}
	\caption{Example of $T(k,z)$ functions, and their redshift-independence for a variety of example models. We consider two baryon interacting models with $\sigma_\mathrm{IDM-b}=10^{-25}$ and $f_\mathrm{IDM-b}=0.5$ \mbox{(IDM-b 1)}, with $\sigma_\mathrm{IDM-b}=5\cdot 10^{-26}$ and $f_\mathrm{IDM-b}=0.1$ \mbox{(IDM-b 2)}; a feebly interacting model with $\Gamma_0=3\cdot10^{-8}\mathrm{Mpc}^{-1}$ and $\Delta N_\mathrm{eff}=0.3$ \mbox{(FIDM)}; and a mixed warm+cold model with $f_\mathrm{wdm}=1.0$ and $m_\mathrm{wdm}=5.3\mathrm{keV}$ \mbox{(Mixed)}\,. Both models that reach a suppression of more than 50\% at $k\sim 10\mathrm{Mpc}^{-1}$ are barely excluded from their likelihood, while the other two lie in the allowed region.
	\label{fig:Tk_example}}
\end{figure}

We can therefore model the suppression at the level of the linear matter power spectrum as a redshift-independent factor
\begin{equation}\label{eq:supdef}
	T^2(k) = \frac{P^\mathrm{model}_\mathrm{lin}(k,z)}{P^{\Lambda\mathrm{CDM}}_\mathrm{lin}(k,z)}~,
\end{equation}
where the redshift dependence on the right hand side roughly cancels out for the models under consideration in this work. We show this behaviour in \cref{fig:Tk_example} for several example models, chosen to show the broad range of suppressions we study in \cref{sec:results}.

For the particular case of thermally produced WDM this can be parametrised~\cite{Bode:2000gq,Viel:2005qj} as 
\begin{equation}
T(k)_\rr{WDM} = \left[1+(\alpha k)^{2\nu}\right]^{-5/\nu} \, ,\label{eq:WDMsup}
\end{equation}
where $\nu = 1.12$ and $\alpha$ is the breaking scale that determines where the suppression begins. 

To cover other models in which all of the DM is non-standard,  Ref.~\cite{Murgia:2017lwo} generalised this shape as
\begin{equation}
T(k) = \left[1+(\alpha k)^{\beta}\right]^{\gamma} \, ,
\end{equation}
where $\alpha$ once again specifies the scale of the suppression, $\beta$ roughly controls the shape of the suppression at low wavenumber~$k$, and $\gamma$ corresponds to the shape of the suppression at high wavenumber~$k$.
This parametrisation offers the advantage that models with different underlying theoretical motivations which produce the same suppression described by $\{ \alpha,\beta,\gamma\}$ lead to the same observed flux power spectrum. Therefore, instead of performing a hydrodynamical simulation for every underlying NSDM model, simulations can be performed for well-chosen combinations of $\{ \alpha,\beta,\gamma\}$ in order to build a grid of simulations that can be interpolated for a given model, as was done in Ref.~\cite{Murgia:2018now}.

\enlargethispage*{1\baselineskip}
This $\{ \alpha,\beta,\gamma\}$-parametrisation provided the basis of the \textsc{MontePython} likelihood developed in Ref.~\cite{Archidiacono:2019wdp}, whereby an interpolation is performed on a large grid of pre-computed simulations, as discussed more in detail below. While this parametrisation covers a wide range of suppressions, it can only account for models which result in a complete suppression of the matter power spectrum at large $k\gtrsim 20/\mathrm{Mpc}$. As such, models with a fraction of NSDM cannot be covered. 

Here we extend on this parametrisation by introducing a new parameter, $\delta$:
\begin{equation}
T(k) = (1-\delta) \left[1+(\alpha k)^{\beta}\right]^{\gamma}+\delta \, ,\label{eq:abgd}
\end{equation}
where $\delta$ will give us the overall height of a plateau in the transfer function for large $k$. As such, this parametrisation can be applied to many more models, including mixed cold and non-standard DM or fractions of interacting DM smaller than unity. This $\{\alpha, \beta, \gamma, \delta\}$-parametrisation covers most models which do not have relevant interactions at small redshift, with a suppression of the power spectrum that does not exhibit strong oscillations. The topic of oscillations is discussed below in \cref{ssec:oscillations}.

With this new formalism, we have developed a new {\sc MontePython} Lyman-$\alpha$ likelihood, which we call {\tt Lyman-ABGD}\,\footnote{This likelihood will be made publicly available upon publication of this paper.}. However, since the additional parameter space is increasingly difficult to cover extensively, our method of interpolating the simulations has been improved with respect to the one used in Ref.~\cite{Archidiacono:2019wdp}, as discussed in \cref{ssec:interpolation}.

\newpage
\subsection{Simulations}

Given the added dimension of variations in the $\delta$ parameter, in principle a new grid of simulations would have to be built and the old one would have to be discarded. Naturally, such a procedure is not particularly computationally efficient. Instead, we augmented the previous $\{\alpha,\beta,\gamma\}$-grid by additional simulations that are designed to cover the added parameter space. Furthermore, since the suppression is modelled as independent of the underlying cosmological parameters, we have investigated two separate grids of simulations: the \textit{cosmo/astro simulations}, designed to capture the effects of varying the $\Lambda$CDM parameters, and the \textit{suppression simulations}, designed to capture the additional effects caused by the NSDM (similar to the approach of Ref.~\cite{Murgia:2018now}).

As described in Ref. \cite{Murgia:2018now}, for the \textit{cosmo/astro simulations} we have used \texttt{GADGET-3}~\cite{Springel:2005mi} (which includes smoothed particle hydrodynamics), with a box length of $20$Mpc/$h$ and $768^3$ gas and $768^3$ CDM particles. This allows for a theoretical resolution of roughly $0.16h$/Mpc up to around $121h$/Mpc. With these simulations, we have considered separately the variations of 5 values of $\sigma_8 \in [0.754\ ..\ 0.904]$ and 5 values of $n_\mathrm{eff} \in [{-2.3474}\ ..\ {-2.2674}]$. Here $n_\mathrm{eff}$ is the slope of the matter power spectrum at a scale of $0.009$s/km, which is typical for the Lyman-$\alpha$ forest. Furthermore, we have considered three values of the instantaneous reionization redshift $z_\mathrm{reio} \in [7,9,15]$ and three values for the UV fluctuation amplitude $f_\mathrm{UV} \in [0,0.5,1]$. This gives us $13$ grid points. In addition to these actual grid points, we have also extended (by linear extrapolation) the grid to cover 21 values of $\sigma_8 \in [0.5..1.5]$ and 15 values of $n_\mathrm{eff} \in [{-2.6}\ ..\ {-2.0}]$ to catch the most extreme models, giving us a total of $39$ points. However, as observed in \cref{sec:results}, many of these additional parameter regions are not actually sampled during any of our runs.
Regarding the impact of inhomogeneous (patchy) reionization, we do not implement in this work the corrections to the 1D flux power quantified and discussed in Ref.~\cite{molaro21}, and we will devote to a future work a full template fitting analysis with this effect included. A preliminary investigation shows that including these effects will not bias our final results in terms of dark matter constraints.

Additionally, we have varied the thermal equation of state of the IGM ($T = T_0 (1+\delta)^{\gamma_T-1}$) by considering three mean temperatures at redshift $z = 4.2$ of $T_0 \in [6\,000,9\,200,12\,500]\mathrm{K}$ and three power law indices $\gamma_T \in [0.88,1.24,1.47]$, giving a $3\times 3$ grid of different IGM thermal histories. This initial grid has been extended (by linear extrapolation) with $8$ additional values of the temperature in $T_0 \in [600..17\,500]$ ($8 \times 3$ points), and $6$ additional values of the slope $\gamma_T \in [0.704..1.694]$ ($6 \times 3$ points), for a total of $8+24+18=50$ additional thermal state grid points. This means that in total we have $8+13=21$ true variations of the above parameters, and the total number of extended points is around $89$.

Finally, we have post-processed each of these $21$ simulations (or $89$ extrapolations) with $17$ different values of the desired mean flux, as further described in Ref.~\cite{Murgia:2018now}. This gives us a total of $21 \times 17 = 357$ simulation points (or a total of $89 \times 17 = 1513$ extrapolations) in the cosmo/astro parameter space. For a given run we assume a power-law behaviour for $T_0(z) = T_0^a [(1+z)/(1+4.2)]^{T_0^s}$ and $\gamma_T(z) = \gamma_T^a [(1+z)/(1+4.2)]^{\gamma_T^s}$, and for each redshift we interpolate the thermal history within the precomputed grid. We also allow for a different mean flux for each of the four redshifts, $\bar{F}(z)$.

\newpage
Our \textit{suppression simulations} have all been run with $512^3$ particles in the same $20$Mpc/h box, with the inputs of the simulations generated by a modified version of \texttt{2LPTic}. These simulations can be classified into three categories.
\begin{enumerate}
	\item First, we use several additional simulations for the thermal suppression shapes of the type of \cref{eq:WDMsup} from Ref.~\cite{Archidiacono:2019wdp} for different values of the corresponding parameter \mbox{$\alpha \in [0.0043,0.0227]\,$Mpc/$h$}, corresponding to WDM masses of $[2,3,4,5,6,7,8,9]\,$keV.
	\item Second, to extend to non-WDM cases, we use 109 simulations for different combinations of the $\{\alpha,\beta,\gamma\}$ parameters.
	\item Third, we have added 83 new simulations that have been initialized with the generalised transfer function of \cref{eq:abgd}. However, to save computational costs, in this case we have explicitly fixed $\gamma=-1.5\beta$\,. This particular choice best captures the usual shape of suppression of most of the considered interacting DM models.
\end{enumerate} 
A list of all the corresponding suppression parameters of the total of 200 suppression simulations according to \cref{eq:abgd} can be found in \cref{tab:abgdvalues} of \cref{sec:tests}.

\subsection{The interpolation}\label{ssec:interpolation}

To interpolate in this sparse grid built from different kinds of simulations, we had to use new interpolation approaches with respect to those used in Refs.~\cite{Murgia:2018now,Archidiacono:2019wdp}. First and foremost, since the problem is separated in suppression simulations and cosmo/astro simulations, we used different approaches for the two types.

For the cosmo/astro simulations we used the same ordinary Kriging method as in Ref.~\cite{Murgia:2018now}. The Kriging method effectively weighs the grid point simulations using weights that depend on the distance in the cosmo/astro parameter space. We take the values of the cosmological parameters $\{ n_\mathrm{eff}, \sigma_8, z_\mathrm{reio}\}$ for the given model and, together with the astrophysical parameters, we derive the Euclidean distances $d$ in the cosmo/astro parameter space. Then we use the function $1/(d+\epsilon)^\eta$ to compute the unnormalised weights for all grid points. Here $\epsilon = 10^{-6}$ is an arbitrary small offset, and $\eta=6$ is an arbitrary enhancement factor.

For the suppression simulations, the three distinct types of suppression functions forced us to abandon the ordinary Kriging approach (117 of the 200 simulations are at $\delta=0$).\footnote{Indeed, we observed that the ordinary Kriging was biased towards the simulations with $\delta=0$. The distance in $\delta$ could never be bigger than $1$, leading typically to relatively small distances to the large number of $\delta=0$ simulations. By their sheer number, the contribution of $\delta=0$ suppressions was then almost always overestimated.} Instead we devised a more general approach.

\enlargethispage*{2\baselineskip}
\subsubsection*{Beyond Kriging}
The overall idea of the Kriging approach is to efficiently compute optimal weights $w_i$ for each grid point $i$, which can be used to estimate the flux power spectrum of the given model as a weighted sum over the flux power spectra of the individual grid points. While the true weights are only known at the given grid points\footnote{For a suppression equivalent to that of grid point $j$, obviously the choice $w_i = \delta^j_i$ is the correct weighting.}, it is reasonable to assume that an input close to the grid point should also return a flux power spectrum close to the one obtained for the grid point.

The precise way of defining \enquote{close} could lead to given methods such as ordinary Kriging (which assumes the weight is inversely proportional to some power of the distance in parameter space) or Gaussian Process regression (which uses a Kernel function instead to define how close a given set of parameters is). However, in our case the parameter space is sampled especially densely only in a hyperslice ($\delta=0$) and thus a method that depends on the distance within the parameter space itself (as for ordinary Kriging and naive Gaussian process regression) is not necessarily optimal. For this reason, we instead consider the distance directly in the suppression function for a grid of wavenumbers $k$ logarithmically sampled between $0.01\mathrm{h/Mpc}$ and $200\mathrm{h/Mpc}$ with $500$ samples.\footnote{In principle, it would be possible to construct a Gaussian process emulator also in this wavenumber basis. Given the resulting high dimension of the Gaussian process and the corresponding difficulty of construction and testing, we leave this approach to future work.} 

\enlargethispage*{1\baselineskip}
In this basis, it is possible to choose weights that minimise this distance. More explicitly, we want to obtain weights $\vec w$ such that they minimise the mean-squared loss
\begin{equation}\label{eq:loss}
	L(\vec w) = \sum_{j=1}^{N_k} \left\lVert S(\vec w, k_j) - T_\mathrm{model}(k_j)\right\rVert^2~, \qquad \text{where} \qquad S(\vec w, k) = \sum_{i=1}^{200} w_i T_i(k)~.
\end{equation}
The optimal choice of weights in this case could be determined by a simple least-squares fit. However, we have an additional condition on our weights $\vec w$\,; the underlying assumption of the Kriging procedure (independently of how exactly it is done) is that the corresponding suppressions of the flux power spectrum are an accurate estimate of the model flux power spectrum suppression:

\begin{equation}\label{eq:fluxsup_fromgrid}
T^\mathrm{Flux}_\mathrm{model}(k) \approx S^\mathrm{Flux}(\vec w, k) = \sum_{i=1}^{200} w_i T^\mathrm{Flux}_i(k)~.
\end{equation}
As such, we would like to additionally ensure that the $w_i$ allow for a good reconstruction of the flux power spectra. The simple least-squares can easily fail in this case: a typical pathological example would be a case where there is a large cancellation in the weighted sum in the loss of \cref{eq:loss}, such as for example $w_0 = 100$ and $w_1 = -99$. In this case, due to the non-linear response of the flux spectrum to linear variations of the input power spectrum, the corresponding sum of \cref{eq:fluxsup_fromgrid} will usually not be a good approximation, i.e. $S^\mathrm{Flux}(\vec w,k) \not\approx T^\mathrm{Flux}_\mathrm{model}(k)$. To prevent such cases, we have to put additional constraints on the weights.

The main issue is that a in simple least-squares approach, weights larger than unity $|w_i| \gg 1$ and smaller than zero $w_i < 0$ are allowed. A very straightforward fix is to simply bound the least-square to only produce weights in the range $0 < w_i < 1$, and we call this approach BLS (bounded least squares). A slightly less restrictive approach is to allow for some weights to be negative as long as they are small. In this case we impose a $L_2$ Tikhonov regularisation on the least-squares solution in order to find solutions with weights as small as possible\footnote{Due to this penalty, only weights for suppression grid points that are really fundamentally relevant to the least-squares can become significantly non-zero. As such, one does not expect a small weight for many models (which would again lead to bad predictions, as even faraway grid points would contribute), but instead only a few relevant weights to be significantly non-zero. This is also what we observe for a few test cases.}.

\subsubsection*{Mathematical description}
The mathematical description can be explicitly written as follows:
\begin{itemize}
	\item Bounded method:
	we solve the bounded least-squares problem \cite{stark1995bounded}
	$\vec w = \mathrm{argmin}\ L(\vec w)$
	with the bounds $\forall j : 0 < w_j < 1$. The solution to this problem can only be found iteratively.
	\item Regularised method:
	we solve the Tikhonov-regularised least-squares problem \cite{calvetti2003tikhonov,golub1999tikhonov}
	$\vec w = \mathrm{argmin}\ L(\vec w) + \lambda ||\vec{w}||^2$\,, which gives the explicit solution
	\begin{equation}
	w_i = \left[M^T M + \lambda I \right]^{-1}_{ij} M_{lj} \cdot T_\mathrm{model}(k_l)~,
	\end{equation}
	where $M_{ij} = T_j(k_i)$ is the grid model matrix and $I_{ij} = \delta_{ij}$ is the identity matrix. For $\lambda\to0$ this reduces to the ordinary least-squares problem, and otherwise the weights are additionally $L_2$ regularised.
\end{itemize}
We investigate the performance of the two weight determination methods in \cref{sec:tests} for a few specific test cases.

Both cases attempt to minimise the loss of \cref{eq:loss} and should, therefore, give accurate representations of the desired model suppression, i.e. one should find $S(\vec w,k) \simeq T_\mathrm{model}(k)$, at least approximately. In \cref{sec:results} we mention how well this is accomplished for each of the models by providing the maximum values of the deviation over all wavenumbers, i.e.
\begin{equation}\label{eq:selfdistanceweighted}
	m_d(\vec w) = \max_{j} \left\lvert S(\vec w,k_j)^2-T_\mathrm{model}(k_j)^2\right\rvert~.
\end{equation}

\subsubsection*{Step-by-step description}
The approach to find the total flux power spectrum associated to any model is then
\begin{enumerate}
	\item Find the $\Lambda$CDM-equivalent model for the given input model, and use its corresponding parameters $\{\sigma_8, n_\mathrm{eff}, z_\mathrm{reio}\}$ together with the nuisance/astrophysical parameters $\{f_\mathrm{UV}, \gamma_T^s, \gamma_T^a, T_0^s, T_0^a, \bar{F}(z)\}$ in the ordinary Kriging to interpolate in the 357 (or 1513 extended) cosmo/astro simulations to obtain the total $\Lambda$CDM-equivalent flux power spectrum for the model. For this, compute the Euclidean distance $d_i$ of the model parameters to those of the grid point $i$, and set the weights to $\alpha_i = 1/(d_i + \epsilon)^\eta$. Then normalise the weights $\alpha_i = \alpha_i / \sum_j \alpha_j$\,, and compute the unsuppressed flux power spectrum as $P^\mathrm{Flux}_\mathrm{\Lambda{}CDM-equivalent}(k) \approx \sum^{1513}_{i=1} \alpha_i P^\mathrm{Flux}_i(k)$. 
	\item Compute the suppression $T_\mathrm{model}(k)$ for the given input model with the help of \cref{eq:supdef}. Use one of the two interpolation methods above to find the weights $\vec w$ of the 200 suppression simulations and subsequently use \cref{eq:fluxsup_fromgrid} to compute the overall suppression $S(\vec w,k)$ of the flux power compared to the $\Lambda$CDM-equivalent model.
	\item Multiply the suppression of step 2 with the overall $\Lambda$CDM-equivalent flux power spectrum of step 1 to obtain the suppressed flux power spectrum for the given model. The final flux power spectrum is $P^\mathrm{Flux}_\mathrm{model}(k) = P^\mathrm{Flux}_\mathrm{\Lambda{}CDM-equivalent}(k) S(\vec w,k)^2$\,.
\end{enumerate}

\subsection{Additional checks}
Given the complexity of the procedure, many of these steps rely on various assumptions, and as such we have included several additional checks to make sure that the assumptions are not broken in a way that would lead to incorrect constraints.

\enlargethispage*{2\baselineskip}
\subsubsection*{Sanity checks}
First, we always perform two \emph{sanity checks} -- conditions that should always be true for any point we are attempting to compute.
\begin{enumerate}
	\item We check that the two cosmological parameters $\sigma_8$ and $n_\mathrm{eff}$ for the given cosmology can be represented within the grid. We perform the same check also for the thermal parameters $T_0(z)$ and $\gamma(z)$ \footnote{We noticed that in an earlier version of our code, this check was not performed for the thermal parameters, but we explicitly checked that the results agreed nonetheless, demonstrating the negligible dependence of these conclusions on the thermal model.}.
	\item We check that $T^2(k_\mathrm{eq})\approx 1$ to $1\%$ accuracy. Otherwise, the definition of a $\Lambda$CDM equivalent model would become ambiguous.
\end{enumerate}
Note that the same sanity checks were also used in Ref.~\cite{Archidiacono:2019wdp}.

\subsubsection*{Coverage tests}\label{ssec:nodata}
To ensure that the constraints we obtain arise from the underlying model constraints and not from the imposed sanity checks (and thus the fidelity of our grid), we also show what limits we would obtain from running without Lyman-$\alpha$ data altogether, similar to what was done in Ref.~\cite{Archidiacono:2019wdp}. In this case, if these \emph{coverage tests} span the entire parameter space that is being constrained, our results are not biased by these sanity checks. This is the case for most of the runs, as explicitly shown in \cref{fig:n0,fig:nm2,fig:mixed}. For the run in \cref{fig:fidm_initial} we notice that the coverage test showed insufficient coverage, prompting us to investigate this case more closely, see \cref{sec:results}. Note that for the coverage test runs we still employ the Planck dataset (in order to minimise required runtime), such that the \emph{coverage test} will never extend beyond the Planck constraints. 

\subsubsection*{Derived parameters}
Beyond these sanity checks, we also record several values for each point that can optionally be used in post-processing to further assess the consistency of the results.
We record the $m_d(\vec w)$ of \cref{eq:selfdistanceweighted}, the sum of the two largest and the four largest weights, and the \emph{area criterion} of Ref.~\cite{Murgia:2017lwo},
\begin{equation}
	\delta A = 1-\frac{\int_{k_\mathrm{min}}^{k_\mathrm{max}} P_\mathrm{1D}(k)/P^{\Lambda\mathrm{CDM}}_\mathrm{1D}(k) \mathrm{d}k}{(k_\mathrm{max}-k_\mathrm{min})}~,
\end{equation}
with the usual definition of
\begin{equation}
	P_\mathrm{1D}(k) = \frac{1}{2\pi} \int k P(k) \mathrm{d}k~.
\end{equation}
We can use these to check if either the $m_d(\vec w)$ is larger than expected (this is mostly not the case, see \cref{sec:results}), if the final result has a substantial contribution from two or four weights, or if the area criterion yields similar results to those expected from Ref.~\cite{Murgia:2017lwo}. We note, however, that the area criterion was calibrated solely for the case of fully NSDM models. Instead, in the cases of mixed cold and non-standard DM, the area criterion no longer gives a direct estimate of the suppression\footnote{For full fraction, the area criterion directly correlates with the mean scale of suppression, and can therefore be used for constraints. In cases where $f \neq 1$, a certain value of the area criterion can either correspond to an early and shallow or a late and deep suppression, both of which are differently disfavoured by the data.}, and so it cannot be used to directly obtain bounds on these models.

\subsubsection*{Test cases}
In addition to these checks that we perform for every single model that we analyse, we also tested our likelihood explicitly for a few test cases presented in \cref{sec:tests}.  First, we checked that leaving out a single grid model and attempting to predict it with all other grid models yields favourable results. Furthermore, we explicitly ran two new simulations for the $n_b=0$ model of DM baryon scattering, chosen to be close to the best-fit region we find in our analysis, as discussed in detail in \cref{ssec:testcases}.

\subsection{Oscillations in the original function}
\label{ssec:oscillations}
In principle it could be assumed that any oscillations in the original suppression function would have to be modeled very explicitly. In practice, however, due to the non-linear mixing of scales, small oscillations of the linear power spectrum are typically \enquote{washed out}, meaning that the predictions for the total flux power spectrum agree between that of a linear power spectrum with oscillations and a linear power spectrum where the oscillations are explicitly smoothed out, as tested explicitly in Refs.~\cite{Murgia:2018now, Archidiacono:2019wdp}). A more detailed and quantitative investigation of models with weak and strong oscillations is left for future work.

%% file: Models.tex
\section{Non-standard dark matter models}
\label{sec:models}

In this section we aim to summarise the theoretical foundations of the various models for which we have derived constraints using the newly-developed {\tt Lyman-ABGD}. As these models are thoroughly described in the literature, here we highlight only the key aspects of each model, and refer to previous work for more dedicated discussions.
We first discuss `standard' thermal WDM, described in \cref{sec:mixed}, where we also consider the case of mixed WDM+CDM scenarios. We then consider a fraction of DM interacting with baryons (hydrogen) in \cref{sec:dmb}, where we focus on two different temperature scalings of the cross section. Finally, in \cref{sec:fidm} we describe a model where the DM interacts feebly with an additional relativistic component of the dark sector (dark radiation).

\newpage
\subsection{Mixed warm and cold dark matter}
\label{sec:mixed}
Thermally produced WDM is well-known to suppress the formation of structures on small scales, and therefore to suppress also the matter power spectrum. For this, the scale on which the power is suppressed significantly is tightly related to the WDM mass. As such, structure formation observations allow to place stringent bounds on the mass of a thermal relic. Indeed, such observations have allowed to exclude WDM masses much smaller than $\mathcal{O}$(keV) -- under the assumption of all the DM being warm~\cite{Viel:2005qj, Viel:2013fqw, Palanque-Delabrouille:2019iyz, Garzilli:2019qki} -- and therefore rule out several relevant DM candidates, such as non-resonantly-produced sterile neutrinos, which are formally equivalent to WDM and are also bounded by X-ray analyses. On the other hand, models in which we have a mixture of two DM components -- one warm and one cold -- induce a much shallower suppression of structure growth, and therefore the bounds in these mixed scenarios can be relaxed~\cite{Boyarsky:2008xj,Baur:2017stq}.

In this work we consider a pure thermally-produced WDM component of an energy density fraction $f_\mathrm{WDM}$ and with mass $m_\mathrm{WDM}$\,, which will usually coexist with a CDM component. One can compute the corresponding temperature of the thermal distribution function as
\begin{equation}
	T \approx T_{1\nu} \cdot \left(\omega_\mathrm{DM} f_\mathrm{WDM} \cdot \frac{94.13 \mathrm{eV}}{m_\mathrm{WDM}}\right)^{1/3}~,
\end{equation}
where $T_\mathrm{1\nu}$ is the temperature of a light relic (such as a neutrino).
Note that there is a re-mapping of such a thermal WDM model to one of sterile Dodelson-Widrow neutrinos produced by non-resonant oscillations in the early universe with a quasi-thermal distribution function. 

In particular, we use Refs.~\cite{Colombi:1995ze,Boyarsky:2008xj, Bozek:2015bdo} to convert from the thermal mass $M$ to the Dodelson-Widrow mass $m_s$
\begin{equation}
	\frac{m_s}{3.90\mathrm{keV}} = \left(\frac{M}{\mathrm{keV}}\right)^{1.294} \left(\frac{0.25 \cdot 0.7^2}{\Omega_M h^2}\right)~,
\end{equation}
where $\Omega_M$ is the energy density fraction of the thermal WDM component.

We note that for the mixed warm+cold DM models we consider that the mass is always so large $\mathcal{O}$(keV) such that we do not need to worry about the redshift-dependence of the suppression function, since the species has already become non-relativistic much before radiation-matter equality.

\subsection{Dark matter - baryon interactions}
\label{sec:dmb}
Models in which the DM can scatter with baryons (either hydrogen, helium, or electrons) have been studied extensively in the literature~\cite{Chen:2002yh,Boehm:2004th,Sigurdson:2004zp,Melchiorri:2007sq,ArkaniHamed:2008qn,Dvorkin:2013cea,Ali-Haimoud:2015pwa,Munoz:2015bca,Kadota:2016tqq,Munoz:2017qpy,Gluscevic:2017ywp,Boddy:2018kfv,Ali-Haimoud:2018dvo,Barkana:2018lgd,Boddy:2018wzy,Bringmann:2018cvk,Emken:2018run,Slatyer:2018aqg,Xu:2018efh,Maamari:2020aqz,Ali-Haimoud:2021lka,Buen-Abad:2021mvc,Nguyen:2021cnb, Rogers:2021byl}, due in part to their complementarity with direct detection experiments and their ability to address some of the existing cosmological tensions. As such, we only summarise the main aspects of such interactions, and we point to the aforementioned references for more in-depth discussions. 

Here we will follow the notation and conventions of Ref.~\cite{Becker:2020hzj}. As such, we focus on DM interacting with protons via a momentum transfer cross section $\sigma_\mathrm{T}$, which has a power-law scaling  with the DM-baryon relative velocity $v$ as $\sigma_\mathrm{T} = \sigma_{\mathrm{DM}-\mathrm{b}}v^{n_b}$\,. In this work we only consider the physically-justified values $n_b \in \left\lbrace -2, 0\right\rbrace$.
The case of $n_b = -2$ occurs in models with DM dipole moments~\cite{Sigurdson:2004zp}, while $n_b = 0$ corresponds to contact interactions~\cite{Chen:2002yh}, as probed in direct detection experiments. 
Note that while the case of $n_b=-4$ arises in models of DM with a fractional electric charge (or millicharge)~\cite{Melchiorri:2007sq}, and is, therefore, of great interest to current cosmology, these interactions have a stronger impact on the CMB than on \mbox{Lyman-$\alpha$} data. Indeed, as already shown in Refs.~\cite{Dvorkin:2013cea, Xu:2018efh, Slatyer:2018aqg}, \lya~data does not improve upon CMB constraints for this model.
Finally, within this framework, we assume that both the DM and the baryons follow a  Maxwell velocity distribution\footnote{For a more general distribution, Ref.~\cite{Ali-Haimoud:2018dvo} extends this using the Fokker-Planck formalism. The implementation of the proposed Beyond-Fokker-Planck approximation is left for future work.} in the early universe, and that these two fluids are non-relativistic. 

The effects of such interactions on the background and thermodynamic quantities are discussed extensively in Refs.~\cite{Slatyer:2018aqg, Becker:2020hzj}. In terms of the matter matter power spectrum, which is the most relevant quantity for \lya~data, the main effect is a suppression of power on small scales due to the added drag between the two species. In \cref{fig:Tk_example} we show an example of the impact these interactions have on the transfer function, for the $n_b=0$ case.

\enlargethispage*{1\baselineskip}
\subsection{Feebly-interacting dark matter}
\label{sec:fidm}
Interactions between DM and an additional relativistic component of the dark sector -- known as dark radiation (DR) -- have long been a source of interest in cosmology~\cite{Archidiacono:2011gq,Diamanti:2012tg,Cyr-Racine:2013fsa,Chu:2014lja,Rossi:2014nea,Buen-Abad:2015ova,Lesgourgues:2015wza,Cyr-Racine:2015ihg,Schewtschenko:2015rno,Krall:2017xcw,Archidiacono:2017slj,Buen-Abad:2017gxg,Archidiacono:2019wdp, Becker:2020hzj}. Here we will focus on the scenario in which the interactions are temperature-independent, thus impacting both early- and late-time cosmological observables. These scenarios, in which there are contributing effects of both the DM-DR interactions and the additional degrees of freedom from the DR, can alleviate both the $S_8$ and $H_0$ tensions, as shown in Refs.~\cite{Lesgourgues:2015wza, Buen-Abad:2017gxg, Archidiacono:2019wdp, Becker:2020hzj,Schoneberg:2021qvd}.

We parametrise our interaction rate in the same way as in Ref.~\cite{Becker:2020hzj} with $n_\mathrm{DR} = 0$. This means that we can write $\Gamma_\mathrm{DM-DR} = \Gamma^0_\mathrm{DM-DR} (1+z)$ -- which follows from $\Gamma_\mathrm{DR-DM} \propto T^{n_\mathrm{DR}}$.
Additionally, we parametrise the amount of DR by its contribution to $N_\mathrm{eff}$ as $N_\mathrm{DR}$\,. 

As was shown in Ref.~\cite{Buen-Abad:2017gxg}, this model can have two main modes: one in which all of the DM is very weakly interacting with the DR, and one in which only a fraction of the DM is strongly coupled to the DR, forming a so-called \emph{dark plasma}. Here we focus on the former case, which we will henceforth call the \emph{feebly interacting dark matter} (FIDM). We note that this corresponds to the weakly interacting scenario of Ref.~\cite{Buen-Abad:2017gxg} and to the models considered in Refs.~\cite{Lesgourgues:2015wza, Archidiacono:2019wdp, Becker:2020hzj,Schoneberg:2021qvd}.
For a summary of the effect of such interactions on the power spectrum, see Ref.~\cite{Archidiacono:2019wdp}.

%% file: Results.tex
\section{Results}
\label{sec:results}
We use the modified version of \class~\cite{Blas:2011rf} presented in~\cite{Becker:2020hzj} (which features DM interactions), as well as the most up-to-date version of {\sc MontePython}~\cite{Audren:2012vy,Brinckmann:2018cvx} with our implementation of the \texttt{Lyman-ABGD} likelihood, as described in \cref{sec:lya}.

\enlargethispage*{2\baselineskip}
As our baseline dataset, we use the Planck 2018 legacy release including temperature, polarisation, and CMB lensing (corresponding to the \textit{high-$\ell$ TTTEEE}, \textit{low-$\ell$ TT}, \textit{low-$\ell$ EE}, and \textit{lensing} likelihoods)~\cite{Aghanim:2018eyx}; and BAO data from BOSS DR12 \cite{Alam:2016hwk}, the SDSS main galaxy sample \cite{Ross:2014qpa}, the 6dFGS sample \cite{Beutler:2011hx}, the QSO clustering from DR14 eBOSS release~\cite{Ata:2017dya},  Lyman-$\alpha$ forest autocorrelation \cite{Agathe:2019vsu}, and the cross correlation of Lyman-$\alpha$ and QSO \cite{Blomqvist:2019rah}, as done in~\cite{Schoneberg:2019wmt,Becker:2020hzj}. We refer to this baseline combination henceforth as \enquote{Planck+BAO}.

We always sample with flat prior distributions in the cosmological parameters of $\Lambda$CDM --~explicitly~$\{\Omega_b h^2, \Omega_c h^2, h, \ln(10^{10} A_s), n_s, \tau_\mathrm{reio}\}$~, as done in~\cite{Aghanim:2018eyx}. We list the considered models (presented in \cref{sec:models}) and their additional parameters and priors below.
\begin{enumerate}
\item A mixture of thermally produced WDM and CDM (\cref{sec:mixed}). We sample in the mass $m_\mathrm{WDM}$ with a flat prior between 0 and 20 keV and in the abundance $\omega_\mathrm{WDM} = \omega_\mathrm{DM}\cdot f_\mathrm{WDM}$\,. This formulation puts the thermal WDM on the same footing as the CDM, but sampling in $f_\mathrm{WDM}$ would not change the results (as they are driven by a sharp cut-off rather than a skewed distribution).
\item A fraction of DM scattering with baryons with $n_b=0$ (\cref{sec:dmb}). In this case we vary the fraction $f_{\mathrm{DM}-\mathrm{b}}$ with a log-prior $\log_{10}(f_{\mathrm{DM}-\mathrm{b}}) \in [-2,0]$ and the interaction strength $\sigma_{\mathrm{DM}-\mathrm{b}}$ also with a log-prior $\log_{10}(\sigma_{\mathrm{DM}-\mathrm{b}}/\mathrm{cm}^2) \in [-29,-22]$. We fix the DM mass to 1\,GeV.
\item A fraction of DM scattering with baryons with $n_b=-2$ (\cref{sec:dmb}). In this case we vary the fraction $f_{\mathrm{DM}-\mathrm{b}}$ with a log-prior $\log_{10}(f_{\mathrm{DM}-\mathrm{b}}) \in [-2,0]$ and the interaction strength $\sigma_{\mathrm{DM}-\mathrm{b}}$ also with a log-prior $\log_{10}(\sigma_{\mathrm{DM}-\mathrm{b}}/\mathrm{cm}^2) \in [-35,-30]$. We again fix the DM mass to 1\,GeV.
\enlargethispage*{4\baselineskip}
\item Feebly interacting DM (\cref{sec:fidm}). In this case we sample the  interaction strength $\Gamma^0_{\mathrm{DM}-\mathrm{DR}}$ and the amount of dark radiation $N_\mathrm{DR}$, for which we put a lower bound of 0. For the interaction strength we put an upper bound of $\Gamma^0_\mathrm{DM-DR} < 10^{-7} \mathrm{Mpc}^{-1}$, as in Refs. \cite{Becker:2020hzj,Schoneberg:2021qvd}, to avoid the sampling issues associated with the extremely elongated tail of high interaction strength and almost negligible dark radiation content. Note that we do not impose a prior on $N_\mathrm{DR} > 0.07$ as in Refs. \cite{Lesgourgues:2015wza,Archidiacono:2019wdp}, since this would bias our analysis, as discussed in Ref. \cite{Becker:2020hzj}. As before, we consider a DM mass of 1\,GeV.
\end{enumerate}

For each of these models we have performed four separate sampling runs with different combinations of data: \textit{Planck + BAO, Planck + BAO + coverage test, Planck + BAO + Lyman-$\alpha$ MIKE/HIRES (regularised), and Planck + BAO + Lyman-$\alpha$ MIKE/HIRES (bounded).} The idea of the coverage test runs is described further in \cref{ssec:nodata}.

\subsubsection*{Mixture of thermal warm dark matter and cold dark matter}

\begin{figure}[t]
	\includegraphics[width=0.59\textwidth]{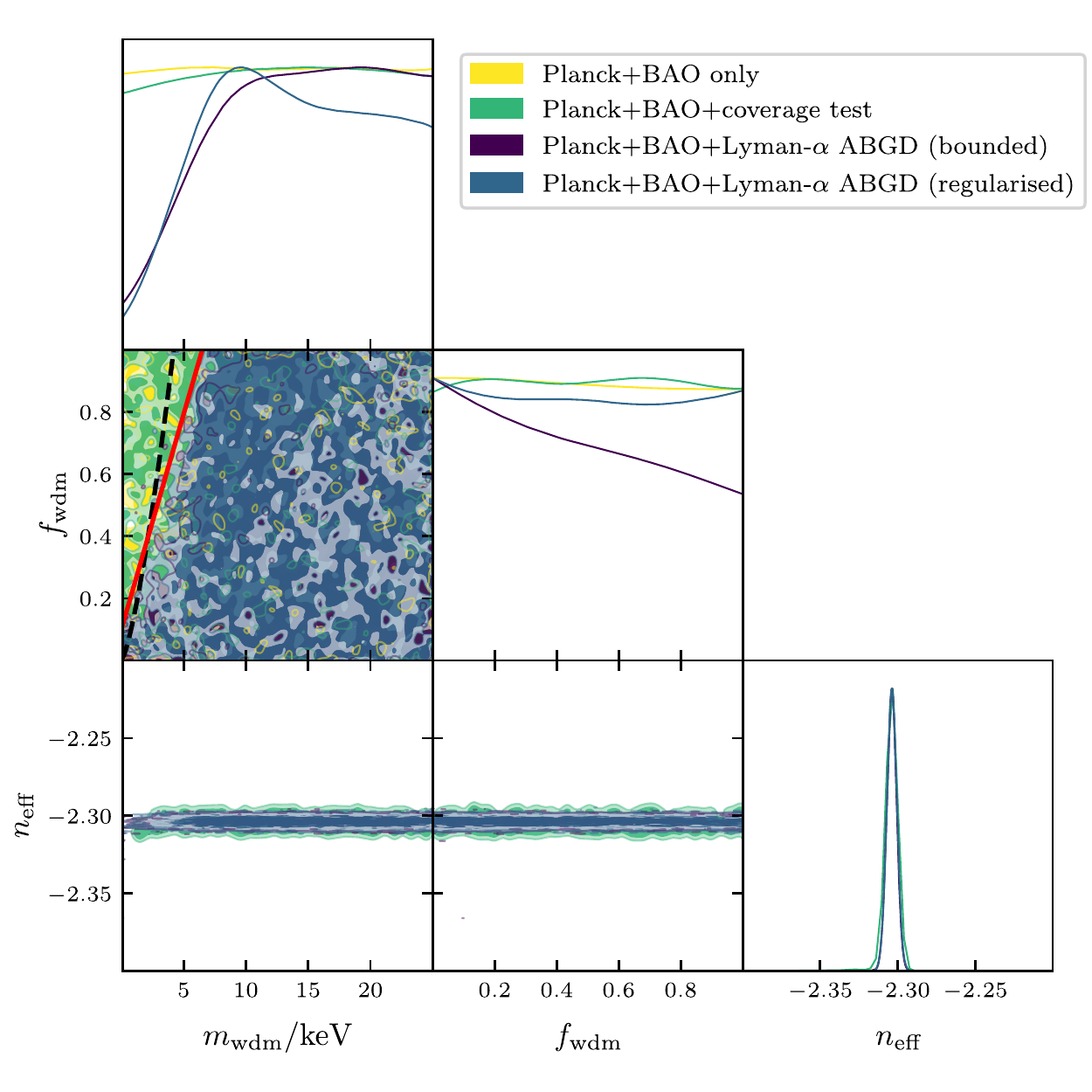}
	\includegraphics[width=0.4\textwidth]{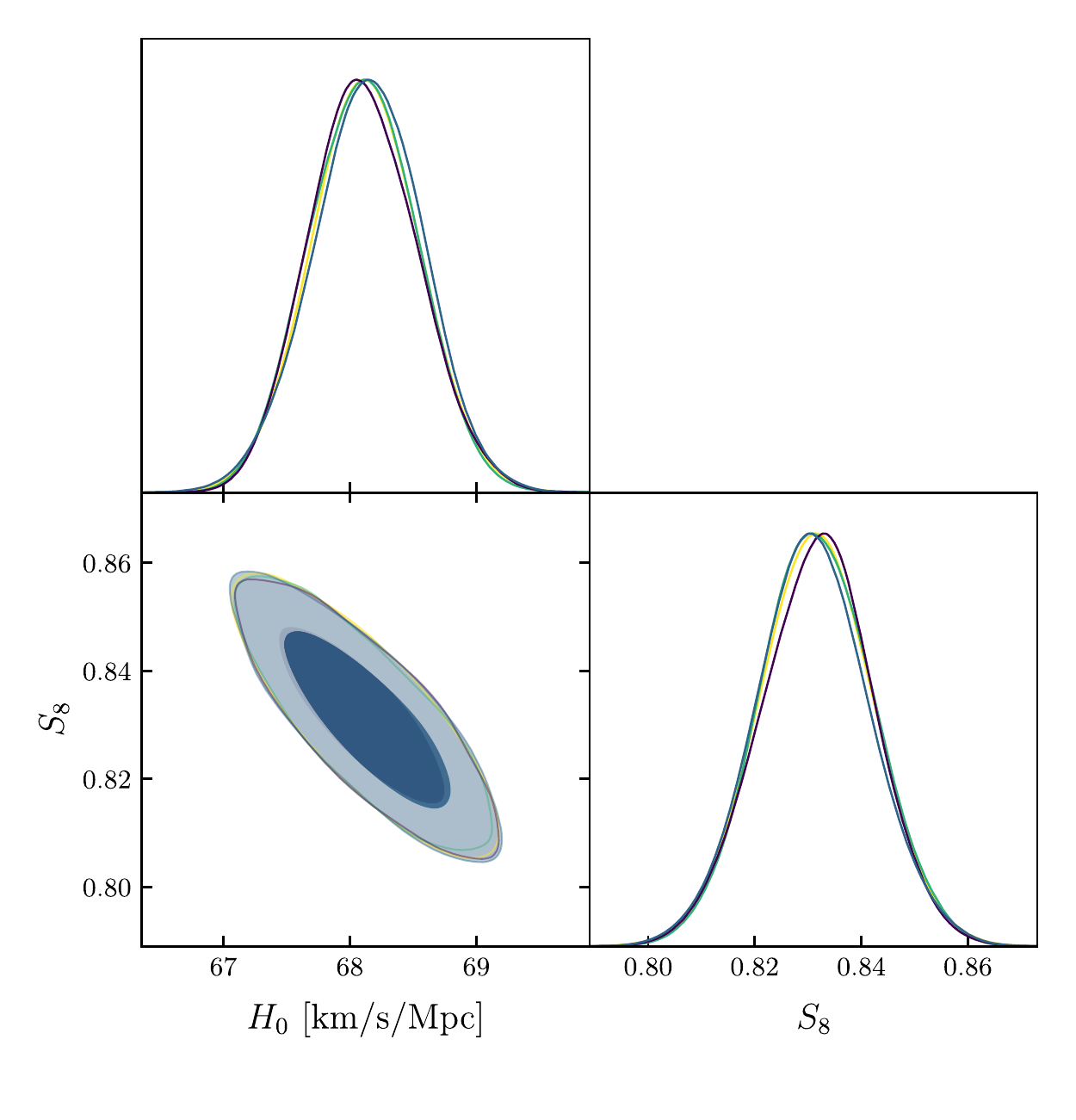}
	\caption{1D and 2D posterior distributions for the key parameters involved in the mixed WDM+CDM scenario; the fraction $f_\mathrm{WDM}$ and the mass $m_\mathrm{WDM}$ as well as the \lya~slope $n_\mathrm{eff}$ (\textbf{left}), and $S_8$ and $H_0$ (\textbf{right}). In red we show our fit of $m_\mathrm{WDM} \gtrsim 7.2\mathrm{keV} (f_\mathrm{WDM}-0.1) $, while in black dashed we show the fit from Ref. \cite{Baur:2017stq}.\label{fig:mixed}}
\end{figure}

Our results for the mixed WDM+CDM run are shown in \cref{fig:mixed}. The coverage test (shown in green) shows that we cover the full region of parameter space sampled, and as such our results are data-driven. As expected, in the Planck case we do not obtain bounds on the WDM mass in the considered range. When including the \lya~data, we obtain a lower bound on the WDM mass, with the bound decreasing as we go to lower fractions. We notice that the bounded and regularised methods yield similar results. The largest distances from the grid representation, $m_d(\vec w)$, are 9\% for the bounded and 10\% for the regularised method. Requiring a smaller distance (e.g. $m_d(\vec w) < 5\%$) does not impact the region where the actual bounds lie. Instead, only mixed models with masses around 10-15keV deviate more than 5\% and would be cut by such a requirement.

\begin{figure}[t]
	\includegraphics[width=0.59\textwidth]{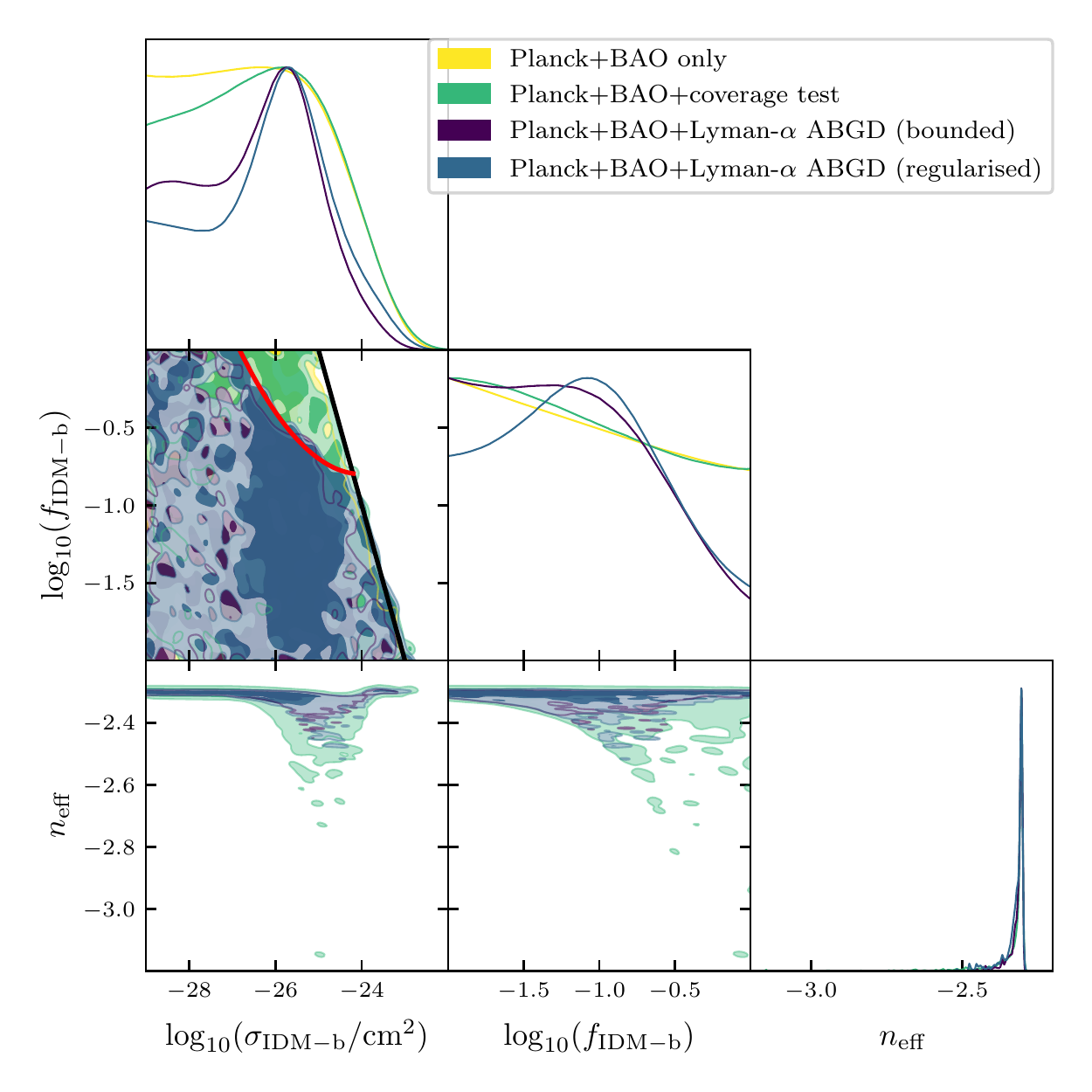}
	\includegraphics[width=0.4\textwidth]{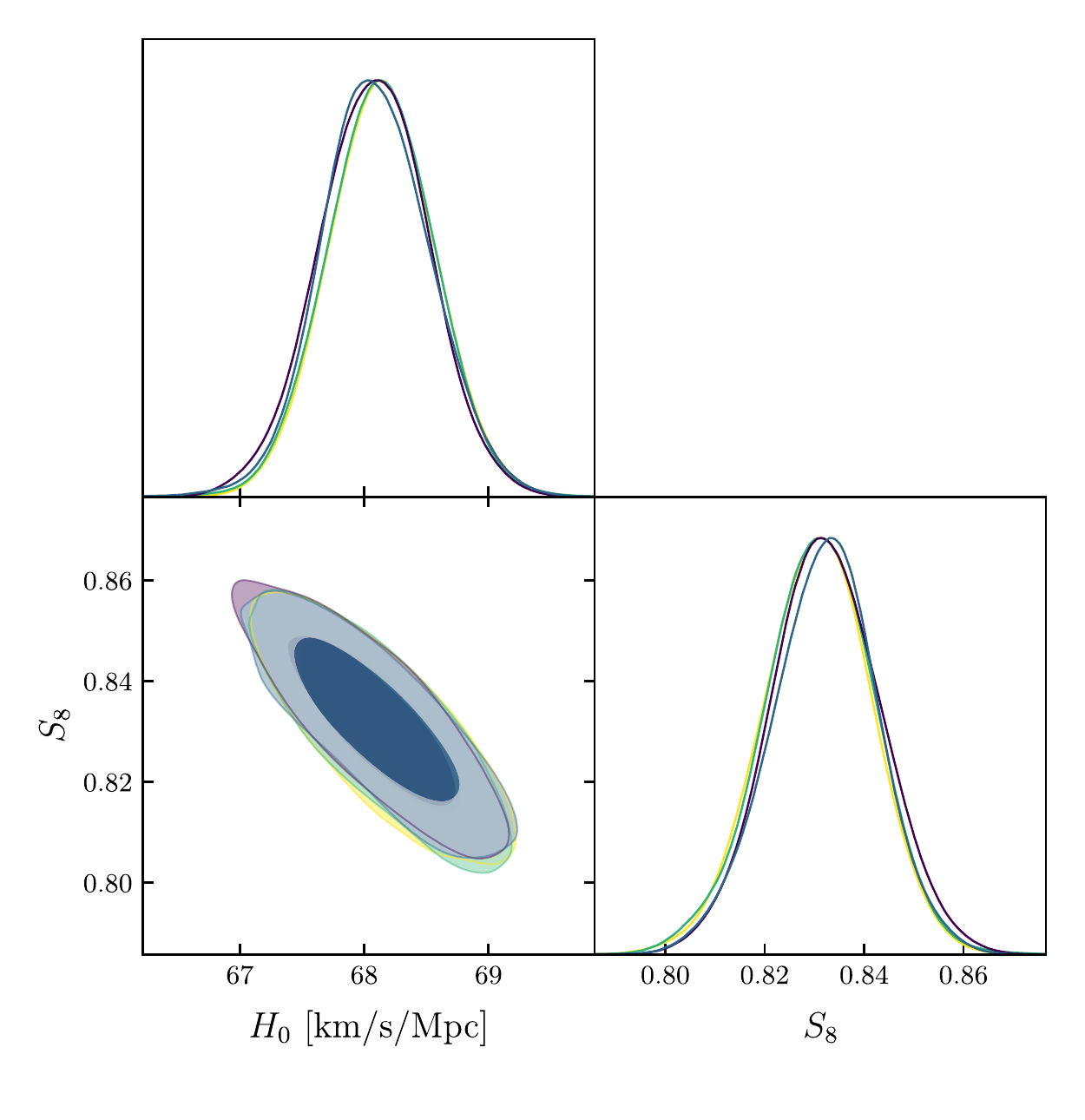}
	\caption{1D and 2D posterior distributions for the key parameters involved in the DM--baryon scattering with $n_b=0$ scenario; the fraction $f_{\mathrm{DM}-\mathrm{b}}$ and the cross section $\sigma_{\mathrm{DM}-\mathrm{b}}$, as well as the \lya~slope $n_\mathrm{eff}$ (\textbf{left}), and $S_8$ and $H_0$ (\textbf{right}). We show the bounds cited in the text as red and black lines. \label{fig:n0}}
\end{figure}

More quantitatively, we find an approximate limit for the mass of the WDM component depending on its fraction as 
\begin{equation} \label{eq:res_mixed}
m_\mathrm{WDM} \gtrsim 7.2\mathrm{keV} (f_\mathrm{WDM}-0.1) \, ,
\end{equation}
shown in \cref{fig:mixed} with a solid red line. 

This should be compared e.g.~to the bound found in Ref.~\cite{Baur:2017stq}, where they obtain $f_\mathrm{WDM} = 0.14 \cdot (1 \mathrm{keV}/m_\mathrm{WDM})^{-1.37}$ when using SDSS, XQ-100, and MIKE/HIRES data (shown in \cref{fig:mixed} with a dashed black line). When fixing $f_\mathrm{wdm}=1$ we find a limit\footnote{Note that taking a limit of a joint bound of two parameters (such as $f_\mathrm{wdm}$ and $m_\mathrm{wdm}$) will usually be more constraining than the bound obtained when \textit{fixing} a priori one of these parameters (in this case, for example, one would get a bound of $6.25\mathrm{keV}$ from the joint-bound limit as $f \to 1$, and we get a bound of $5.9\mathrm{keV}$ in the fixed case). To understand this fact more intuitively, consider a simple two dimensional uncorrelated Gaussian distribution with $-\ln \mathcal{L} \propto x^2+y^2$ and flat priors. The joint bounds on $x$ and $y$ will be given by the circle $x^2+y^2<1$. It is also easy to check that for any fixed $x=\hat{x}$ the likelihood just receives an offset, and thus the one-variable bound on $y$ is always simply $|y|<1$. Instead, the limit as  $x\to x_*$ of the joint bound is $|y| < \sqrt{1-x_*^2}$\,. This simple example demonstrates the general idea that the limit of a joint bound is tighter than the one-variable bound is when fixing the other variables a priori.}
of around 5.9keV. This limit is slightly tighter than that of \cite{Irsic:2017ixq, Viel:2013fqw}. This is because the prior chosen in this work is in $m_\mathrm{WDM}$\,, while the cited literature uses a prior in $1/m_\mathrm{WDM}$\,. The former is motivated from a model-building perspective, while the latter appears in the velocity dispersion that is most directly constrained by the data. Adopting the same prior as \cite{Irsic:2017ixq, Viel:2013fqw} we find $\sim 3\mathrm{keV}$, showing the good agreement of our \texttt{Lyman-ABGD} likelihood with previous results.

\enlargethispage*{1\baselineskip}
Finally we note that neither $S_8$ nor $H_0$ are affected by the presence of (fractional) WDM, as expected given that this model has no impact on the expansion rate of the universe, and in the considered mass range the power spectrum suppression is not on scales that can affect $S_8$. Furthermore, the slope of the linear matter power spectrum on \lya~scales, $n_\mathrm{eff}$ is also unaffected by the presence of WDM, and we recover the standard \lcdm~value.

\subsubsection*{Dark Matter scattering with baryons \texorpdfstring{$\boldsymbol{n_b=0}$}{nb=0}}
In this case, shown in \cref{fig:n0}, we can see that once again the coverage test covers the full region constrained by Planck, and therefore the likelihood can safely be used for this model.
We notice that the constraints from Lyman-$\alpha$ data nicely compliment the Planck constraints, especially where the fraction is relatively high. As the fraction decreases the constraining power relative to the Planck data also decreases, as expected from the non-linear dependence of the flux power spectrum on the suppression caused by the interactions. The biggest deviations from the grid representation $m_d(\vec w)$ in this case are around 4.5\% for the regularised and 5.5\% for the bounded method.

More qualitatively, we can summarise the constraints using linear/quadratic fits as follows:
\begin{eqnarray}
\text{Planck + BAO:} \ & \log_{10}(f_{\mathrm{DM}-\mathrm{b}}) &< -25-\log_{10}({\sigma_{\mathrm{DM}-\mathrm{b}}}/\mathrm{cm}^2) \, , \label{eq:res_nb0_a} \\ 
\text{+ Lyman-$\alpha$ (quadratic fit):} \ & \log_{10}(f_{\mathrm{DM}-\mathrm{b}}) &< -0.8+0.1 \cdot(\log_{10}({\sigma_{\mathrm{DM}-\mathrm{b}}}/\mathrm{cm}^2)+24)^2 \, ,~~~~~~\label{eq:res_nb0_b}\\
\text{+ Lyman-$\alpha$ (linear fit):} \ & \log_{10}(f_{\mathrm{DM}-\mathrm{b}}) &< -0.3 \log_{10}({\sigma_{\mathrm{DM}-\mathrm{b}}}/\mathrm{cm}^2)-8.2 \label{eq:res_nb0_c} \, .
\end{eqnarray}
The constraint from Planck + BAO from eq.~\eqref{eq:res_nb0_a} can also be expressed in a simpler form as $f_\mathrm{IDM-b} \cdot \sigma_{\mathrm{DM}-\mathrm{b}} < 10^{-25}\mathrm{cm}^2$, and is shown in \cref{fig:n0} with a black line. The additional Lyman-$\alpha$ bound obtained when using a quadratic fit from eq.~\eqref{eq:res_nb0_b} is displayed in \cref{fig:n0} with a red line. Finally, the bound obtained if we impose a linear constraint also on the Lyman-$\alpha$ data from eq.~\eqref{eq:res_nb0_c} can be rewritten as $f_{\mathrm{DM}-\mathrm{b}} \cdot \left(\frac{\sigma_{\mathrm{DM}-\mathrm{b}}}{\mathrm{cm}^2}\right)^{0.3} < 6 \cdot 10^{-9}$\,.

In the limit of $f_{\mathrm{DM}-\mathrm{b}}=1$, we find $\log_{10}({\sigma_{\mathrm{DM}-\mathrm{b}}}) < -25.0 $ for the Planck case, in agreement with the results of e.g. Refs.~\cite{Xu:2018efh, Slatyer:2018aqg,Rogers:2021byl}. When adding the \lya~data, this is reduced to $\log_{10}({\sigma_{\mathrm{DM}-\mathrm{b}}}) < -27.4$, which is in very good agreement with the results of Refs.~\cite{Xu:2018efh, Rogers:2021byl} (for a DM mass of 1\,GeV, as considered here). Once again, this agreement for the fully interacting case provides a good check on the likelihood developed here.

\begin{figure}[t]
	\centering
	\includegraphics[width=0.8\textwidth]{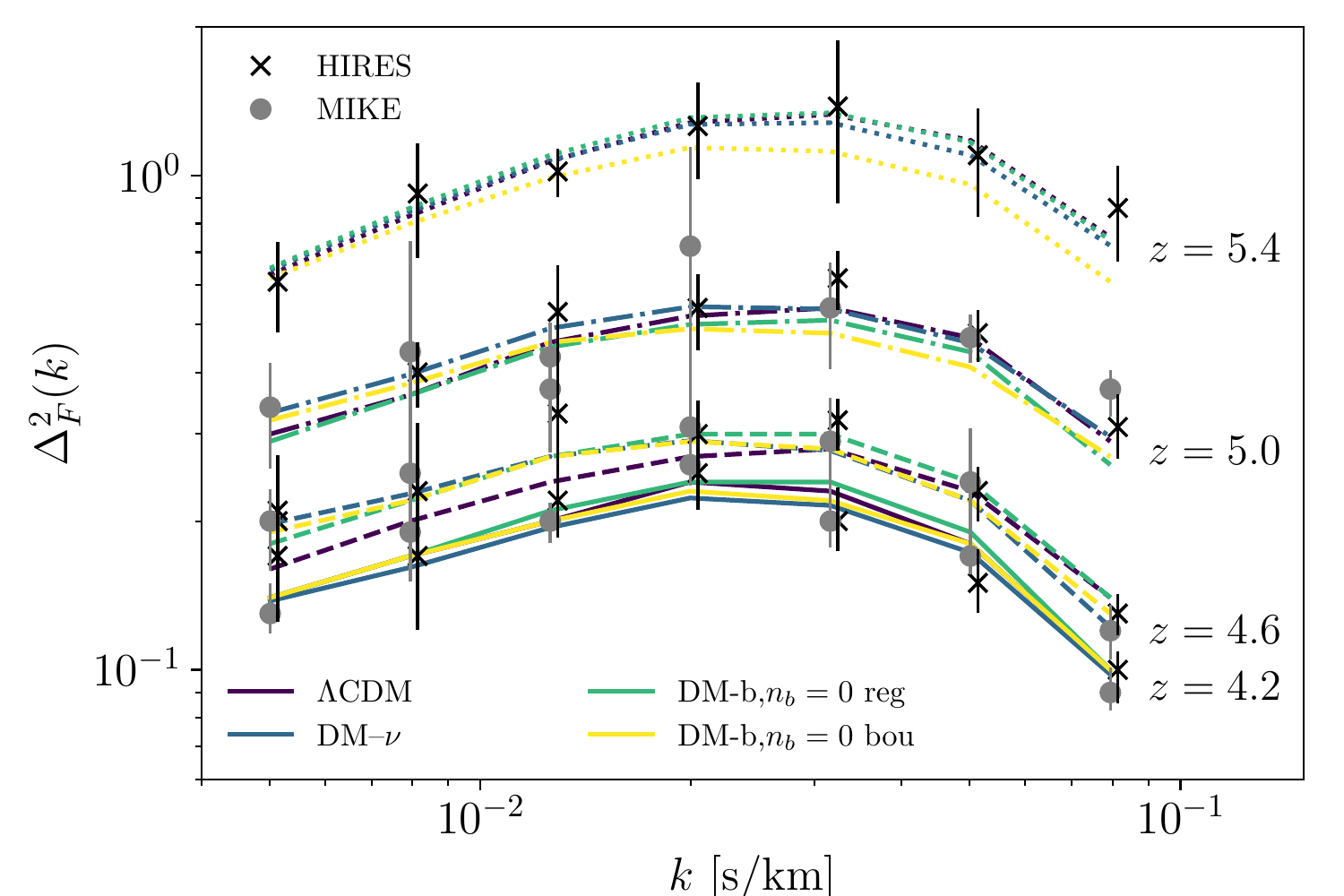}
	\caption{Comparison between the flux for \lcdm~(purple), the bestfit DM--neutrino interacting model from Ref.~\cite{Hooper:2021rjc} (blue), and DM--baryon interactions, for both the regularised (green) and bounded (yellow) methods. The different line styles denote the different redshift bins, while the crosses and dots show the HIRES and MIKE data, respectively. \label{fig:flux_comp}}
\end{figure}

Very interestingly, looking at the 1D posteriors in \cref{fig:n0}, we can see that when including the \lya~data, we find a mild preference for a non-zero interaction strength. This preference is not present in the coverage test, hinting to it being driven by the inclusion of the \lya~data. Indeed, our bestfit model when using the regularised method is $\{\log_{10}({\sigma_{\mathrm{DM}-\mathrm{b}}}) = -25.98, \log_{10}(f_{\mathrm{DM}-\mathrm{b}}) = -1.25, n_\mathrm{eff} = -2.31\}$, while for the bounded case it is $\{\log_{10}({\sigma_{\mathrm{DM}-\mathrm{b}}}) = -26.02, \log_{10}(f_{\mathrm{DM}-\mathrm{b}}) =-0.64, n_\mathrm{eff} = -2.32\}$ .  This preference is reminiscent of the one found recently in Ref.~\cite{Hooper:2021rjc} for DM interacting with massive neutrinos, which was found using the \texttt{Lyman-ABG} likelihood of Ref.~\cite{Archidiacono:2019wdp}. Putting this into numbers, we observe that the regularised fit prefers approximately\footnote{We define in this case the 68.3\% (or 1$\sigma$) CL only in a symmetric way, asking how large the number $L$ has to be such that the range $\mu-L$ to $\mu+L$ with mean $\mu$ contains 68.3\% of samples. We have explicitly checked that other methods (such as water-level filling of a smoothed version of the posterior) give approximately the same numbers, albeit more dependent on the precise smoothing method employed.} $\log_{10}(\sigma_{\mathrm{DM}-\mathrm{b}})  = -26.2 \pm 1.48$ at the 68.3\% symmetric~CL, while the bounded fit prefers $\log_{10}(\sigma_{\mathrm{DM}-\mathrm{b}}) = -26.4\pm 1.51$ at the 68.3\% symmetric~CL. These numbers can give us a \textit{rough} understanding of the preference for non-zero interaction rate.

\newpage
To further understand this mild preference, we show in \cref{fig:flux_comp} the flux we obtain from our bestfit model, as calculated by the \texttt{Lyman-ABGD} likelihood. For comparison, we also show the \lcdm model, as well as the bestfit from Ref.~\cite{Hooper:2021rjc}. As in the latter case, we see that our bestfitting models tilt the overall flux with respect to \lcdm, resulting in a mild enhancement at large scales and a suppression at small scales. 
On the other hand, while Ref.~\cite{Hooper:2021rjc} found a bestfit with a much lower value of $n_\mathrm{eff}$ than the standard value of $n_\mathrm{eff} = -2.3$, here we find a bestfit with only milder departures from this value. However, we can see in the posterior distributions in \cref{fig:n0} that the model can result in lower values of $n_\mathrm{eff}$. In order to test if the likelihood still performs well for these lower values of $n_\mathrm{eff}$, we ran dedicated hydrodynamical simulations for two such cases, as discussed in \cref{ssec:testcases}. As we illustrate there, the results from the simulations and the predictions from the likelihood are in very good agreement. This further points to the mild preference being of physical origin. However, additional tests for this model using compatible \lya~data are left for future work.

\subsubsection*{Dark Matter scattering with baryons \texorpdfstring{$\boldsymbol{n_b=-2}$}{nb=-2}}

\begin{figure}[t]
	\includegraphics[width=0.59\textwidth]{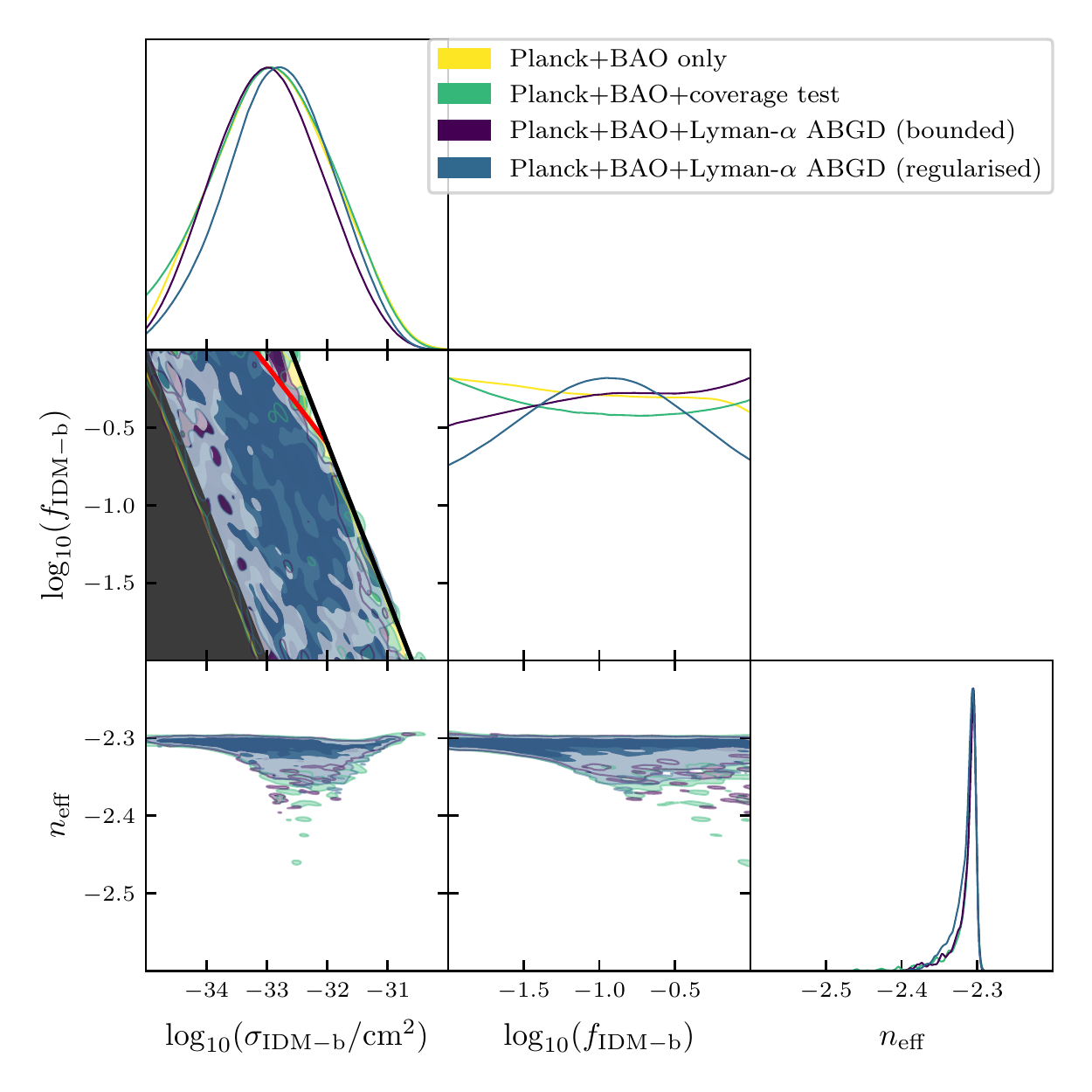}
	\includegraphics[width=0.4\textwidth]{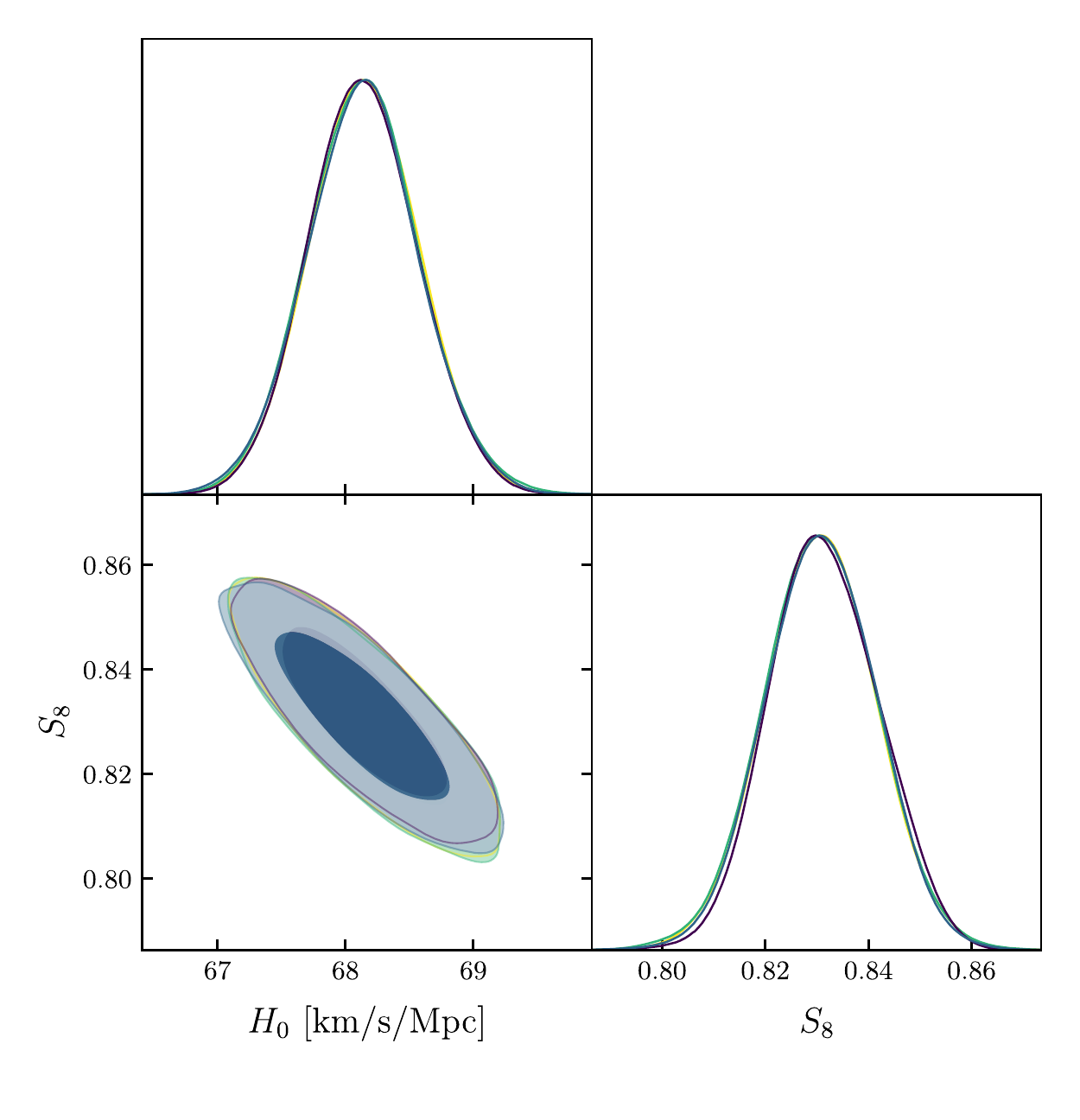}
	\caption{1D and 2D posterior distributions for the key parameters involved in the DM--baryon scattering with $n_b=-2$ scenario; the fraction $f_{\mathrm{DM}-\mathrm{b}}$ and the cross section $\sigma_{\mathrm{DM}-\mathrm{b}}$, as well as the \lya~slope $n_\mathrm{eff}$ (\textbf{left}), and $S_8$ and $H_0$ (\textbf{right}). We show the bounds cited in the text as red and black lines.\label{fig:nm2}}
\end{figure}

This case, shown in \cref{fig:nm2}, is remarkably close to the $n_b=0$ case discussed above; however, there are a few small details that are different in this scenario. First, the \lya~data does not provide much further constraining power than Planck, unlike in the previous scenario. This is expected and a well-known effect in the full fraction case, due to the interactions having more impact at early-times and less impact at late-times \cite{Dvorkin:2013cea}. Furthermore, there are some issues with the computation for very small cross section (as already discussed in Ref.~\cite{Becker:2020hzj}), which are indicted as a blacked-out area and do not impact our results (as shown in \cref{fig:nm2}). However, this means that the 1d posterior can naturally not be fully trusted in this case, since the marginalisation is sensitive to this computational boundary. This also implies that in this scenario the non-zero peak of the 1D posterior of $\log_{10}({\sigma_{\mathrm{DM}-\mathrm{b}}})$ cannot be interpreted as a mild preference for non-zero values, unlike in the $n_b = 0$ case. The biggest deviations $m_d(\vec w)$ in this case are around 1.2\% for the regularised and 0.5\% for the bounded method.

The constraints we derive can be approximately expressed through simple fitted formulae as
\begin{eqnarray}
\text{Planck + BAO:} \ & \log_{10}(f_{\mathrm{DM}-\mathrm{b}}) &< -32.6-\log_{10}({\sigma_{\mathrm{DM}-\mathrm{b}}}/\mathrm{cm}^2) \, , \label{eq:res_nbm2_a} \\ 
\text{+ Lyman-$\alpha$:} \ & \log_{10}(f_{\mathrm{DM}-\mathrm{b}}) &< -16.6-0.5\log_{10}({\sigma_{\mathrm{DM}-\mathrm{b}}}/\mathrm{cm}^2) \, .\label{eq:res_nbm2_b} 
\end{eqnarray}
The bound from Planck + BAO from eq.~\eqref{eq:res_nbm2_a} can also be expressed simply as a constraint on  $f_{\mathrm{DM}-\mathrm{b}} \cdot \sigma_{\mathrm{DM}-\mathrm{b}} < 2.5 \cdot 10^{-33}\mathrm{cm}^2$, and is shown in \cref{fig:nm2} with a black line. The additional bound from the inclusion of Lyman-$\alpha$ data from eq.~\eqref{eq:res_nbm2_b} can be rewritten as \mbox{$f_{\mathrm{DM}-\mathrm{b}} \cdot \sqrt{{\sigma_{\mathrm{DM}-\mathrm{b}}}/\mathrm{cm}^2} < 2.5 \cdot 10^{-17}$} and is illustrated in \cref{fig:nm2} with a red line.

Once again, if we focus on the limit of $f_{\mathrm{DM}-\mathrm{b}}=1$, we find $\log_{10}({\sigma_{\mathrm{DM}-\mathrm{b}}}) < -32.60$ for the Planck case, in close agreement (but slightly more constraining, due to the newer Planck data used here) with the results of e.g. Refs.~\cite{Xu:2018efh, Boddy:2018wzy} (again for a DM mass of 1\,GeV). The further addition of the \lya~data strengthens this bound to $\log_{10}({\sigma_{\mathrm{DM}-\mathrm{b}}}) < -33.20$, which agrees with the results in Ref.~\cite{Xu:2018efh}. As before, this overall agreement shows the reliability of the likelihood in these scenarios.

\subsubsection*{Feebly interacting dark matter}
\begin{figure}[t]
	\includegraphics[width=0.59\textwidth]{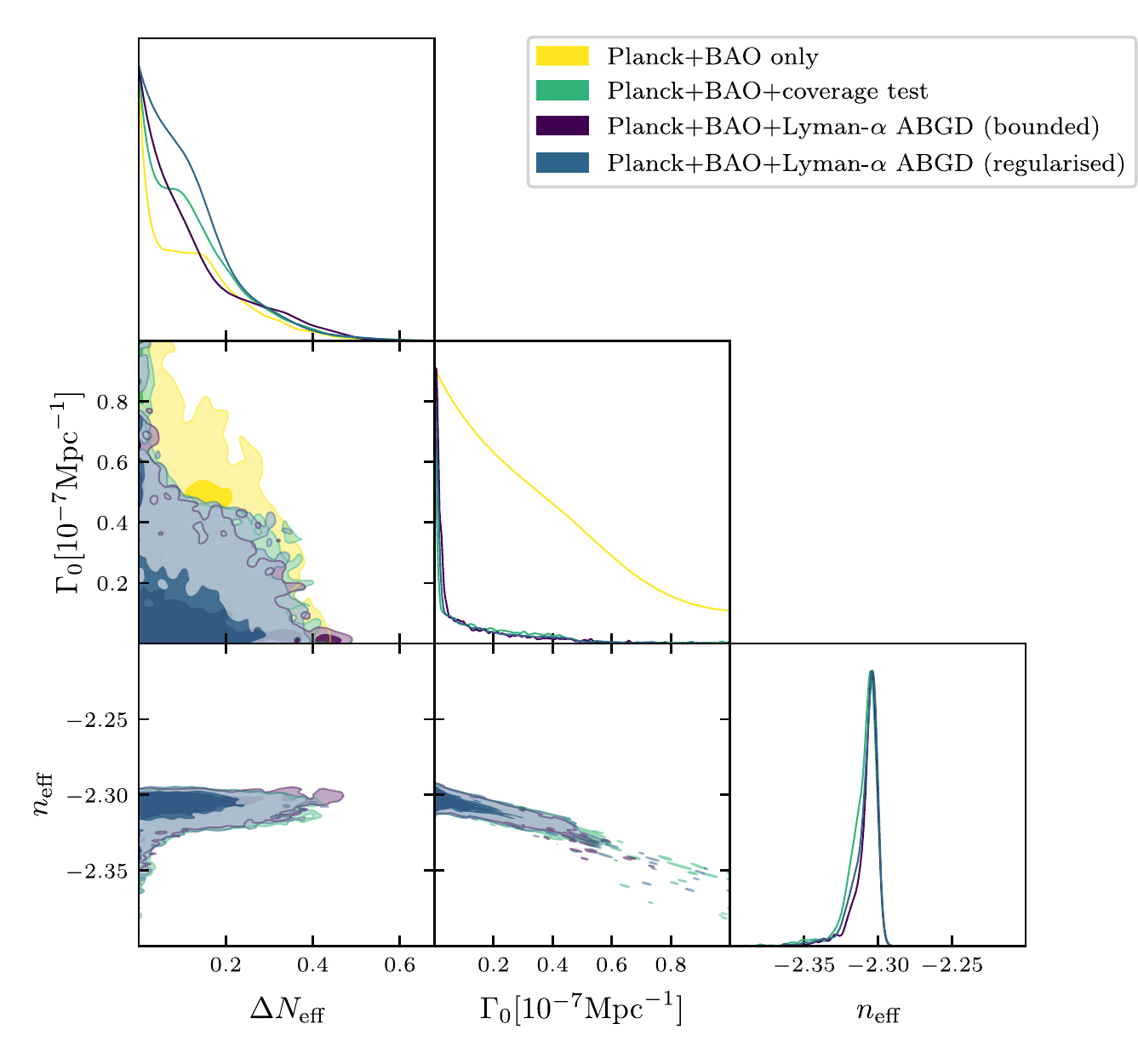}
	\includegraphics[width=0.4\textwidth]{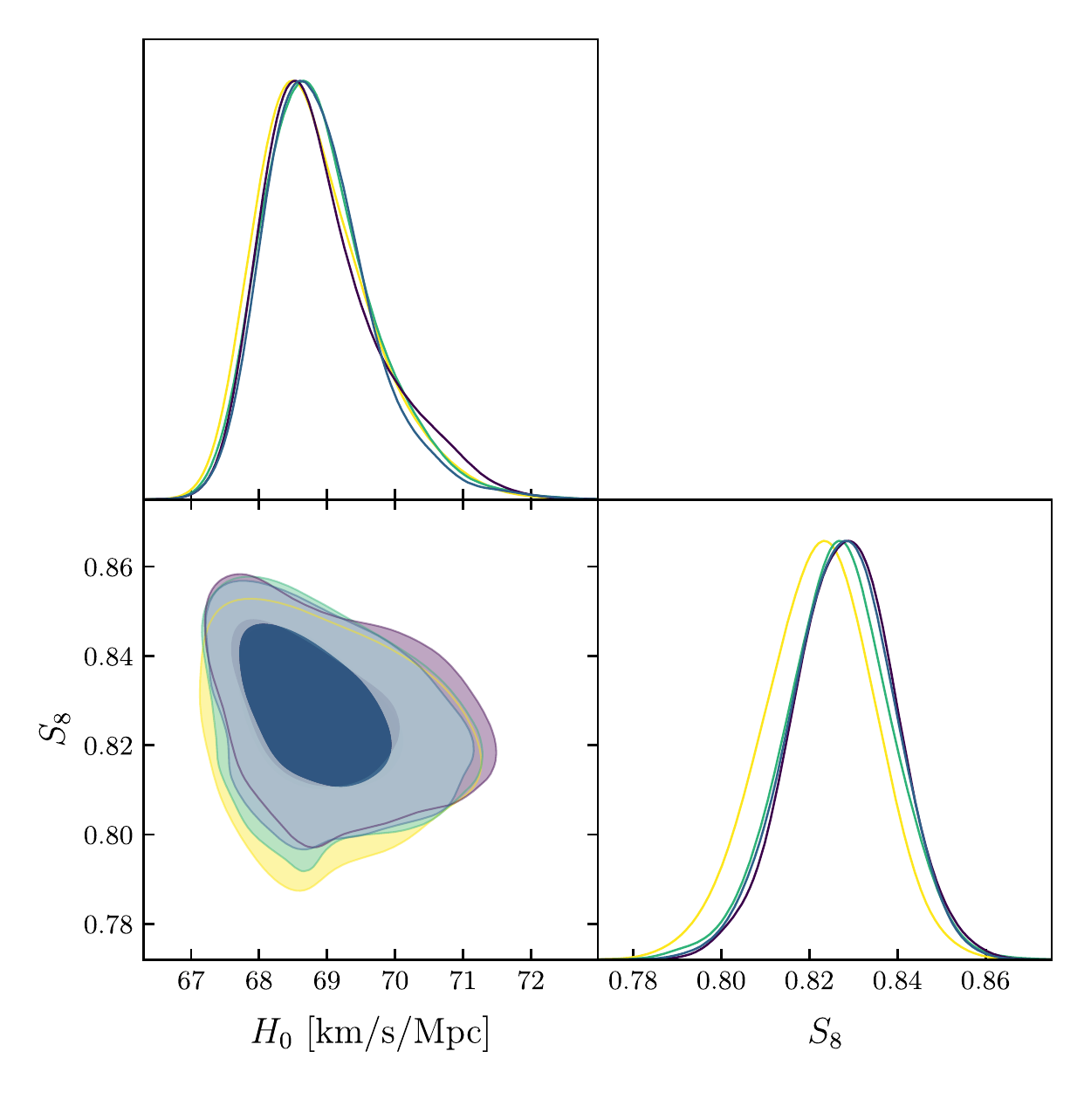}
	\caption{1D and 2D posterior distributions for the key parameters involved in the FIDM scenario; the amount of DR $N_\mathrm{DR} $ and the interaction strength $\Gamma^0_{\mathrm{DM}-\mathrm{DR}}$, as well as the \lya~slope $n_\mathrm{eff}$ (\textbf{left}), and $S_8$ and $H_0$ (\textbf{right}). Note that the units of $10^{-7}\mathrm{Mpc}^{-1}$ for $\Gamma^0_{\mathrm{DM}-\mathrm{DR}}$ correspond to approximately $3\cdot 10^{-4} H_0/h$. \label{fig:fidm_initial}}
\end{figure}

In the initial run of this case, displayed in \cref{fig:fidm_initial}, we noticed that the coverage test was not covering the full parameter space that was covered by Planck. Indeed, in this case the second sanity check of \cref{ssec:interpolation} -- in which we check that $T(k_\mathrm{eq}) \approx 1$ to $1\%$ accuracy -- is not satisfied. Investigating these cases more closely, we noticed that this requirement might be comparatively restrictive, given that many cases with a difference larger than $1\%$ could still be very well represented by the grid. Indeed, the biggest deviations from the representation within the grid $m_d(\vec w)$ in this case are only 1\% for the both regularised and the bounded method. As such, we also show a run where we relaxed this criterion to $5\%$\footnote{This criterion is set as a flag in the likelihood, which can be easily adjusted for each case.}. 

We further note that the constraints from Ref. \cite{Archidiacono:2019wdp} correspond to $\Gamma^0_{\mathrm{DM}-\mathrm{DR}}\cdot 10^7 \mathrm{Mpc} < 0.46$ in our notation (assuming a bosonic DR with 2 degrees of freedom), which exactly corresponds to the bound imposed by the requirement of $T(k_\mathrm{eq}) \approx 1$ to $1\%$ that was already imposed in Ref. \cite{Archidiacono:2019wdp}. We can thus explain \emph{why} the results of Fig. 5 of the reference were saturating their coverage-test (the light-blue contours marked \enquote{Priors}) for the case of $n_\mathrm{DR} = 0$.

\begin{figure}[t]
	\includegraphics[width=0.59\textwidth]{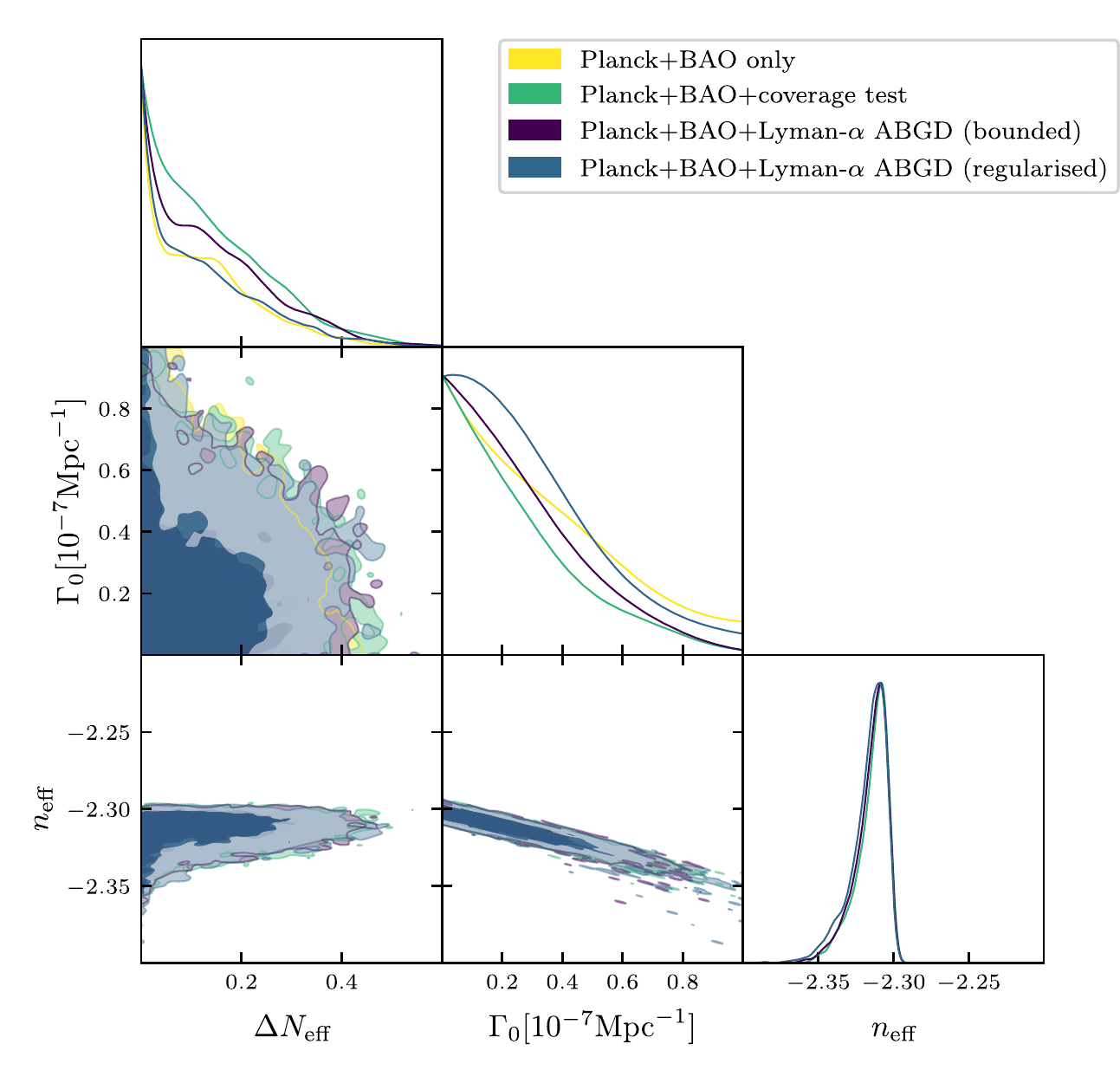}
	\includegraphics[width=0.4\textwidth]{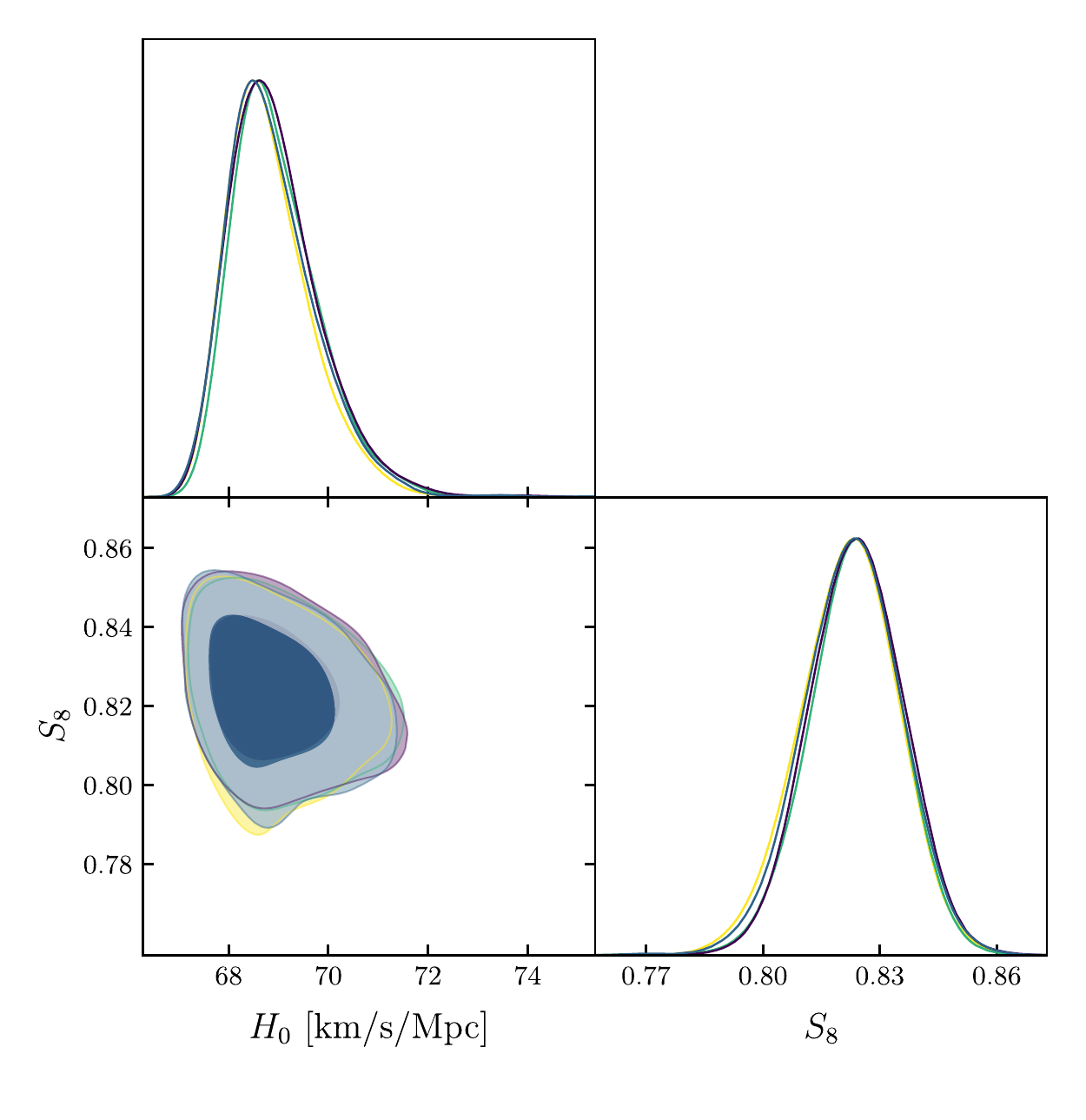}
	\caption{Same as \cref{fig:fidm_initial}, but using a different bound for $T(k_\mathrm{eq})$ as discussed in the text.
	 \label{fig:fidm_corrected}}
\end{figure}

\enlargethispage*{4\baselineskip}
Instead, when adopting the more permissive constraint of $T(k_\mathrm{eq}) \approx 1$ to $5\%$ (shown in \cref{fig:fidm_corrected}), we recover the complete region from Planck with the (extended) coverage test. However, in this case we also notice that the Lyman-$\alpha$ constraints do not further restrict the relevant parameter space compared to the Planck constraints. This can be understood when considering an example suppression such as from \cref{fig:Tk_example} (in that case $\Gamma_0=3\cdot 10^{-8}\mathrm{Mpc}^{-1}$ and $\Delta N_\mathrm{eff}=0.3$), where the very shallow suppression of the FIDM models that are still allowed by Planck data clearly are not sharp or strong enough to cause significant detections with Lyman-$\alpha$ data.

\FloatBarrier

%% file: Conclusions.tex
\section{Conclusions}
\label{sec:conc}

Persistent tensions among different cosmological datasets have spurred interest in models beyond our standard \lcdm~paradigm. One broad class of models in this direction are non-standard dark matter scenarios, in which the DM in the universe has non-negligible velocity or scatterings with other species. These models affect late-time structure formation, resulting in a suppression of the matter power spectrum. As such, small-scale structure data is crucial to understand these scenarios.

Lyman-$\alpha$ data is an invaluable tool for studying hydrogen clouds in the IGM. However, these observations relate to the underlying matter distribution in a highly non-linear way, necessitating dedicated computationally demanding hydrodynamical simulations, which makes parameter scans infeasible. 
To circumvent the need for such simulations, a general approach was proposed in Refs.~\cite{Murgia:2017cvj,Murgia:2017lwo,Murgia:2018now}, which was later turned into a dedicated \textsc{MontePython} likelihood in Ref.~\cite{Archidiacono:2019wdp}. Here we have extended this formalism to encompass a much broader variety of DM models, allowing for suppressions in the matter power spectrum which induce a plateau at large $k$ instead of dropping to zero. This is especially useful for constraining models where a fraction of NSDM coexists with CDM.

Our newly-developed likelihood, dubbed \texttt{Lyman-ABGD}, relies on a grid of 200 hydrodynamical simulations and uses an advanced interpolation scheme to calculate the $\chi^2$ for each sampled model. The likelihood features two different ways of weighting the grid points (bounded and regularised), with a different underlying mathematical description, but leading to similar errors and similar results. Furthermore, the likelihood features several consistency checks, which can be easily toggled by the user. As a proof-of-principle of this approach, we have used the \texttt{Lyman-ABGD} likelihood to constrain several NSDM scenarios.

First, we have revisited the case of WDM, allowing for the WDM to coexist with standard CDM. In this mixed scenario, Lyman-$\alpha$ data allow us to place a lower bound on the mass of the WDM, with the bound moving to higher masses as we allow for smaller fractions. For both the bounded and regularised method, our bound can be summarised as \mbox{$m_\mathrm{WDM} \gtrsim 7.2\mathrm{keV} (f_\mathrm{WDM}-0.1) $}. When moving to the purely WDM case (i.e. when enforcing $f_\mathrm{WDM} =1$), we found a bound of $m_\mathrm{WDM} \gtrsim 5.9\mathrm{keV}$. When instead using a prior of $1/m_\mathrm{WDM}$, as done in the literature, we recover a bound of $\sim 3 \mathrm{keV}$, in good agreement with Refs.~\cite{Irsic:2017ixq,Viel:2013fqw}. This good agreement with previous results serves as a useful cross-check of the likelihood.

We then focused on the case of a fraction of DM scattering with baryons, for two different velocity scalings, $n_b = \{ 0, -2\}$. In both cases we showed that the likelihood is well suited to constrain these models, and both approaches yield the same results. We found an upper bound on the interaction strength that decreases as we go to lower fractions, and we provide analytical results, both in terms of the fraction $\log_{10}(f_{\mathrm{DM}-\mathrm{b}})$ and the cross section $\log_{10}({\sigma_{\mathrm{DM}-\mathrm{b}}}/\mathrm{cm}^2)$, summarised in eqs.~\eqref{eq:res_nb0_a}-\eqref{eq:res_nbm2_b}. To the best of our knowledge, these are the first such bounds obtained from Lyman-$\alpha$ data for these models with varying fractions.  Focusing on the limit of $\log_{10}(f_{\mathrm{DM}-\mathrm{b}}) = 0$, which has been studied previously, we closely reproduced the bounds in the literature when including the Lyman-$\alpha$ data, once again showing the robustness of the method derived here.
\newpage
Furthermore, for the case of a fraction of DM scattering with baryons with a velocity scaling of $n_b =0$, we found a mild preference for non-zero interactions, similar to the preference recently found in Ref.~\cite{Hooper:2021rjc} for DM scattering with massive neutrinos. To understand if this result is driven by our method, we performed dedicated hydrodynamical simulations for two cases close to our best-fit models. In both cases, the flux obtained from the simulations closely matches the reconstructed flux obtained using the interpolation scheme in the likelihood. This shows that the numerical setup of the likelihood itself is unlikely to be the source of the preference, hinting at the existence of a flux that fits the Lyman-$\alpha$ data better than $\Lambda$CDM. This mild preference should, of course, be tested with additional complementary Lyman-$\alpha$ data, as well as a more rigorous investigation of the corresponding thermal histories. We leave these tests to future work.

Finally, we have studied the FIDM scenario, in which all of the DM is feebly coupled to DR, and which has been discussed extensively given its ability to alleviate the $H_0$ and $S_8$ tensions. In our first analysis, we noticed that we could not cover the full region of parameter space, due to our stringent criteria on the deviations from a $\Lambda$CDM-like matter power spectrum at large scales. This shows that the imposed checks are performing as expected. A second analysis, where we relaxed this criterion slightly, shows that we can analyse the full parameter space. However, in this scenario the inclusion of Lyman-$\alpha$ data does not significantly improve on the constraints obtained using Planck+BAO data. The reason for this is that the suppression induced by these models already begins at larger scales, as shown in \cref{fig:Tk_example}, and is, therefore, already tightly constrained by early-universe probes. As such, we conclude that while our likelihood is applicable in this scenario (under more relaxed assumptions), the data itself does not bring any new information.

\enlargethispage*{2\baselineskip}
To conclude, we have justified the need to develop new strategies to use Lyman-$\alpha$ data in order to constrain models beyond the standard $\Lambda$CDM scenario. The likelihood we have developed here, which will be made publicly available, allows for the study of many beyond-$\Lambda$CDM scenarios without needing new computationally-expensive simulations, and could play an important role in constraining scenarios that can alleviate the existing cosmological tensions.

%% file: Appendix.tex
\appendix
\newpage
\section{Testing the performance of the grid}\label{sec:tests}
\subsection*{Leave-one-out tests}\label{ssec:leaveoneouttests}
In these sets of tests, we used the existing grid simulations to check whether our method was working as expected. Instead of using all simulations to fit a given model, we used all simulations except for a single simulation $i$ and tried to predict with these remaining simulations the suppression shape corresponding to the input suppression for simulation $i$. Naturally, this test is not perfect, as it will work better in regions where there are more simulations, and work less well for \enquote{boundary} simulations, which are delivering critical information to the grid that cannot be estimated from the other simulations. These are, for example, simulations with the highest/lowest wavenumber suppressions. Despite these issues, it can be very instructive to show how well a given sampling scheme performs under this criterion.

\begin{figure}
	\centering
	\includegraphics[width=0.49\textwidth]{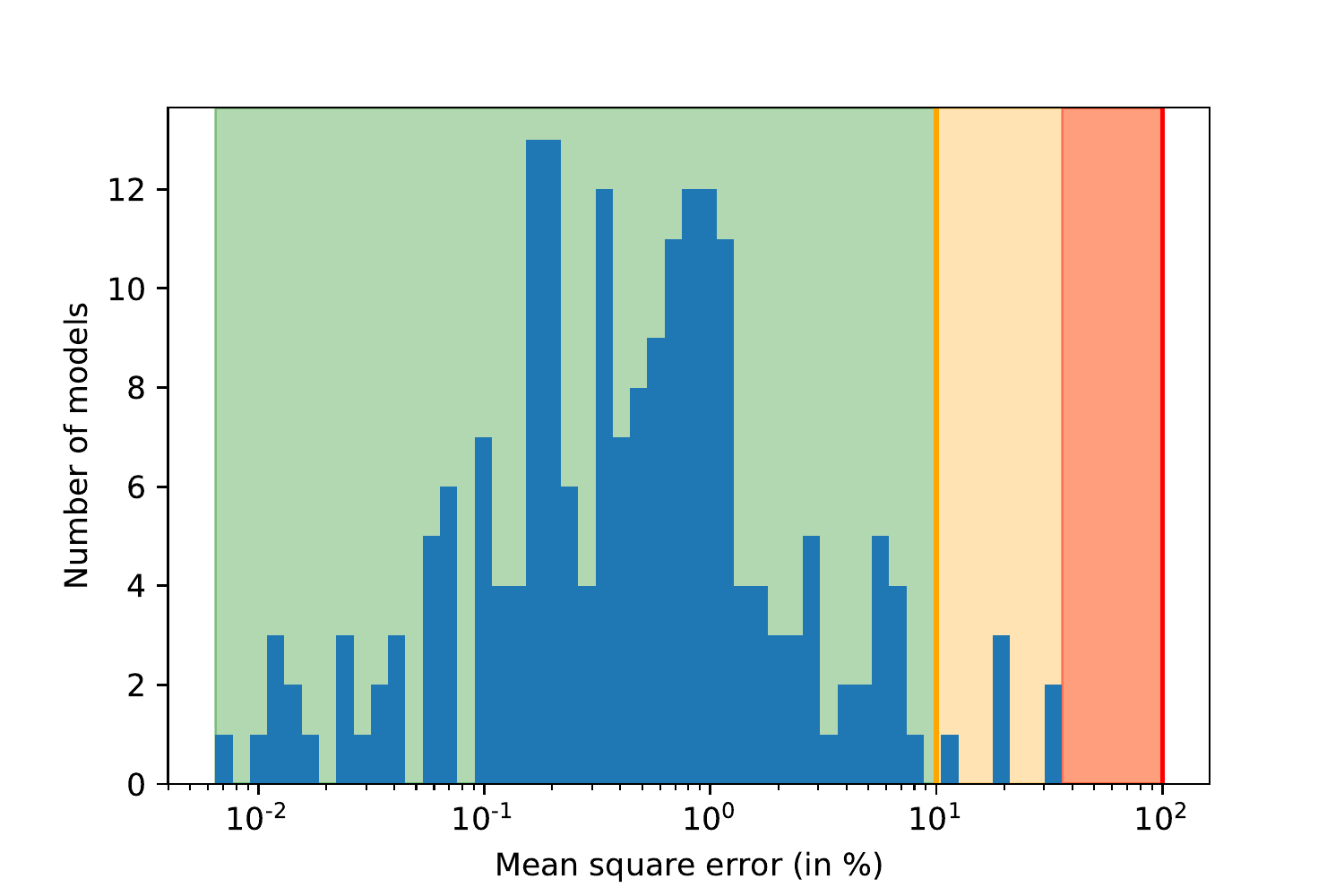}
	\includegraphics[width=0.49\textwidth]{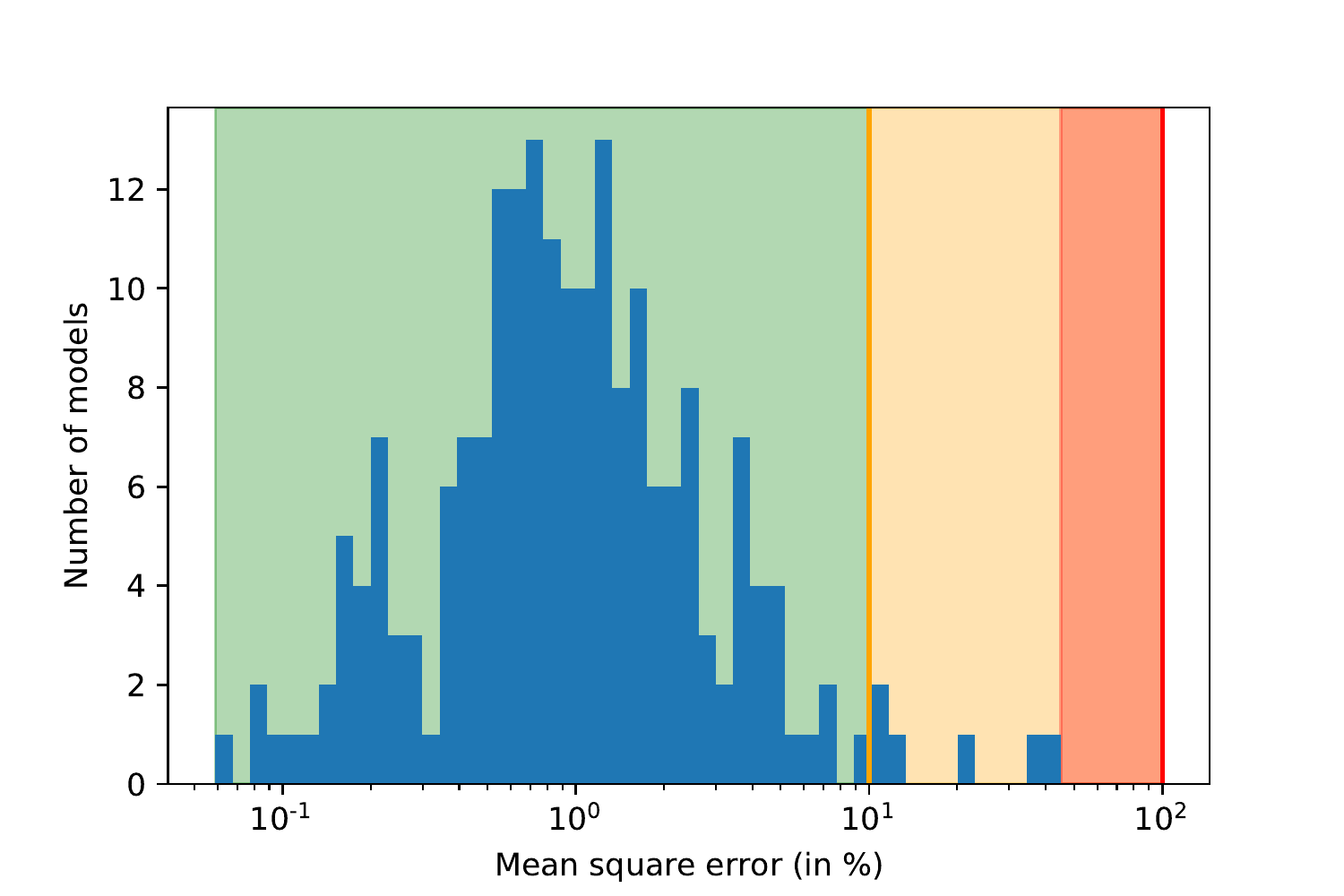}
	\caption{MSQE of leave-one-out tests (see text) for all 200 possible variationsfor the bounded case (\textbf{left}) and regularised case (\textbf{right}). Note that in the regularised case the errors only go down to $\sim 10^{-1}$, rather than $\sim 10^{-2}$ in the bounded case.
	\label{fig:msqe}}
\end{figure}

\enlargethispage*{2\baselineskip}
For this test, we compute the flux power spectrum for model $i$ from the remaining $199$ grid simulations $P^\mathrm{Flux}_{199,i}(k)$, and compare to the flux power spectrum for the $i$-th simulation $P^\mathrm{Flux}_i(k)$ by computing the mean square deviation (MSQE). The histogram of this MSQE for all 200 grid points that can be left out one-by-one is shown in \cref{fig:msqe} for both the bounded case (left) and the regularised case (right). Most of the grid point simulations can be recovered with only a miniscule 1\% mean square error, showing the excellent accuracy of the likelihood. Only around $\sim 10$ of the 200 simulations show errors in recovery bigger than around $10\%$, and are likely to be concentrated at the edges of the grid. We conclude that both methods show excellent grid coverage.

We observe that the regularised case does perform better for almost all simulations, with both the maximum and mean of the distribution skewed towards smaller errors (though the bounded case has a tail towards even lower MSQEs).

\newpage 
\subsection*{On additional simulations}\label{ssec:testcases}
\begin{figure}[t]
	\centering
	\includegraphics[width=0.45\textwidth]{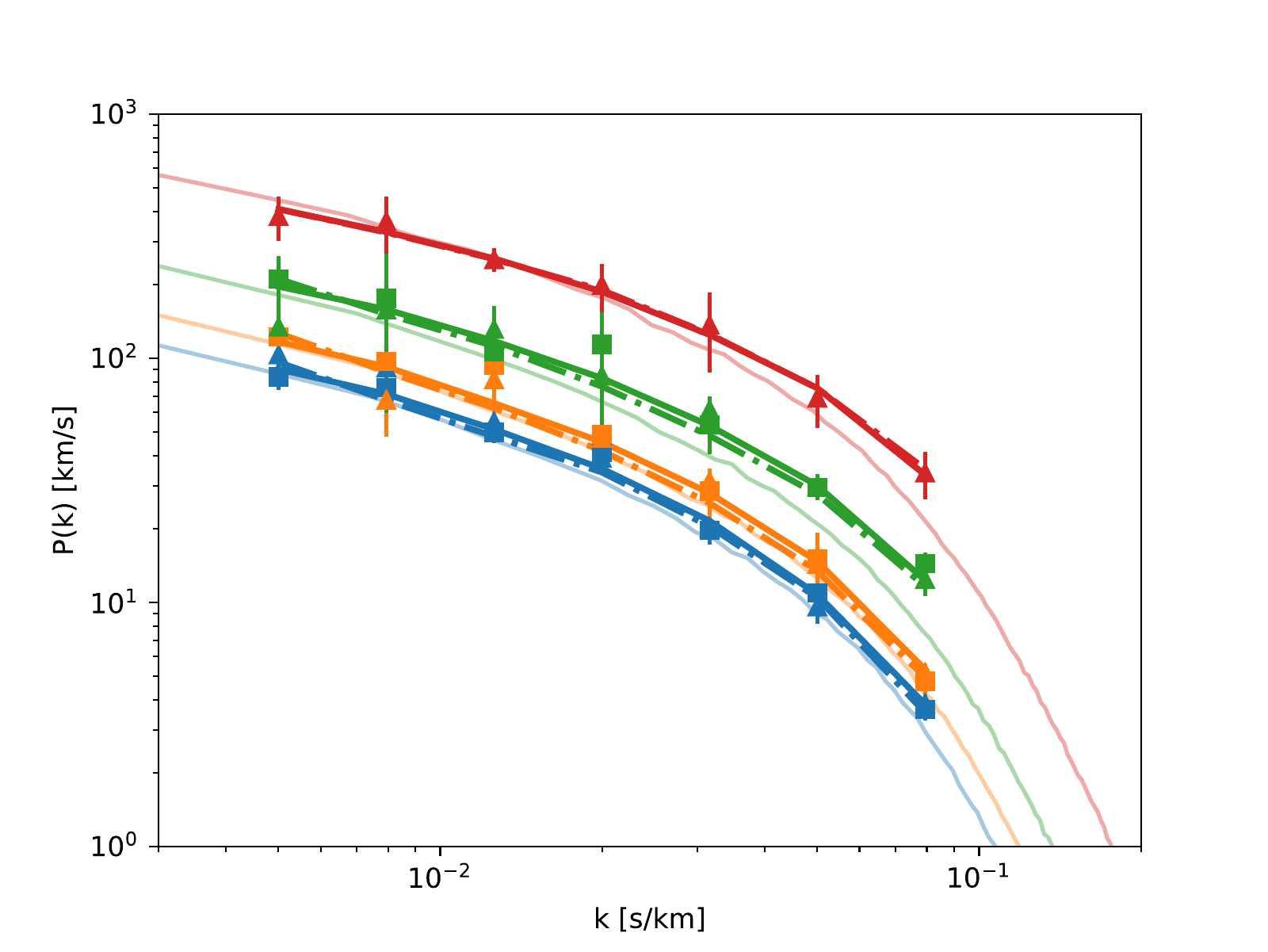}
	\includegraphics[width=0.45\textwidth]{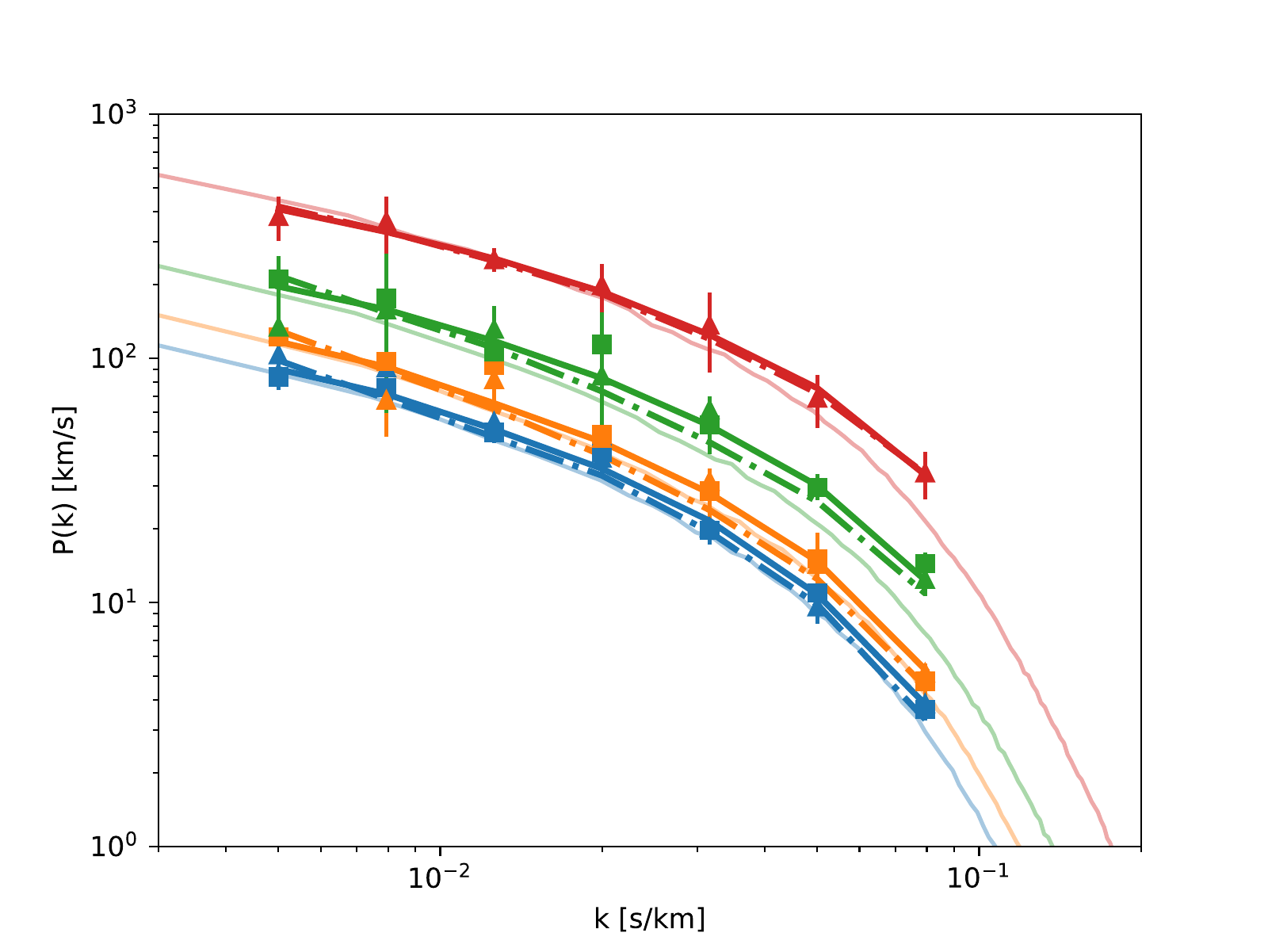}
	\caption{Model with $\sigma_{\mathrm{DM}-\mathrm{b}}= 6.2 \cdot 10^{-26}$ and $f_{\mathrm{DM}-\mathrm{b}}=0.23$. The colors are $[z=4.2 - \mathrm{blue}, z=4.6 - \mathrm{orange}, z=5.0  -\mathrm{green}, z=5.4 - \mathrm{red}]$. The faint lines in the background are the additional simulation spectra themselves, which are then corrected using the astro/kriging grid in solid lines. The dashed lines are instead the predictions from the simulation grid. The triangle correspond to HIRES data, while the squares correspond to MIKE data. \textbf{Left:} Bounded method. \textbf{Right:} Regularised method. \label{fig:compare_boubf}}
	
	\centering
	\includegraphics[width=0.45\textwidth]{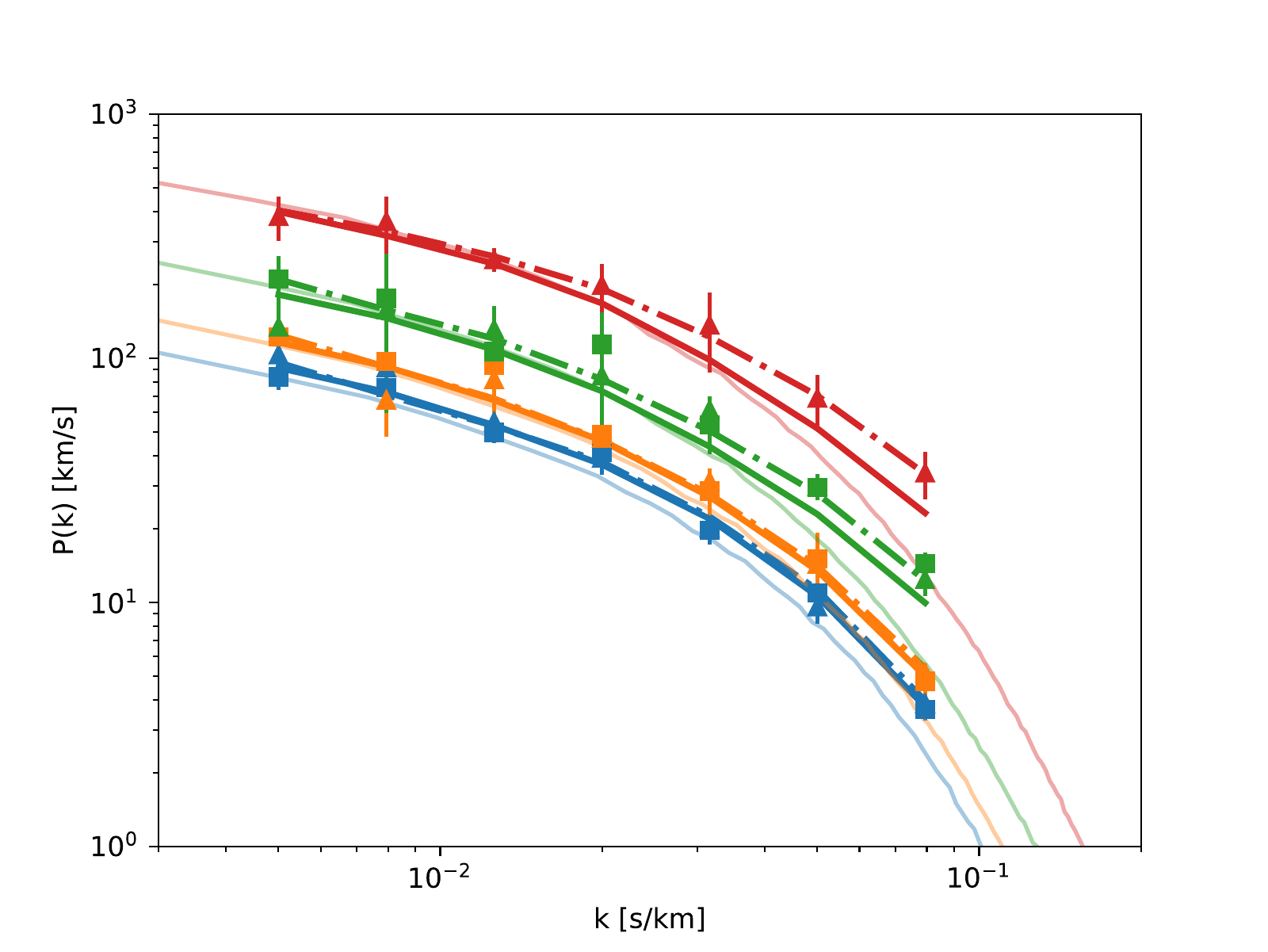}
	\includegraphics[width=0.45\textwidth]{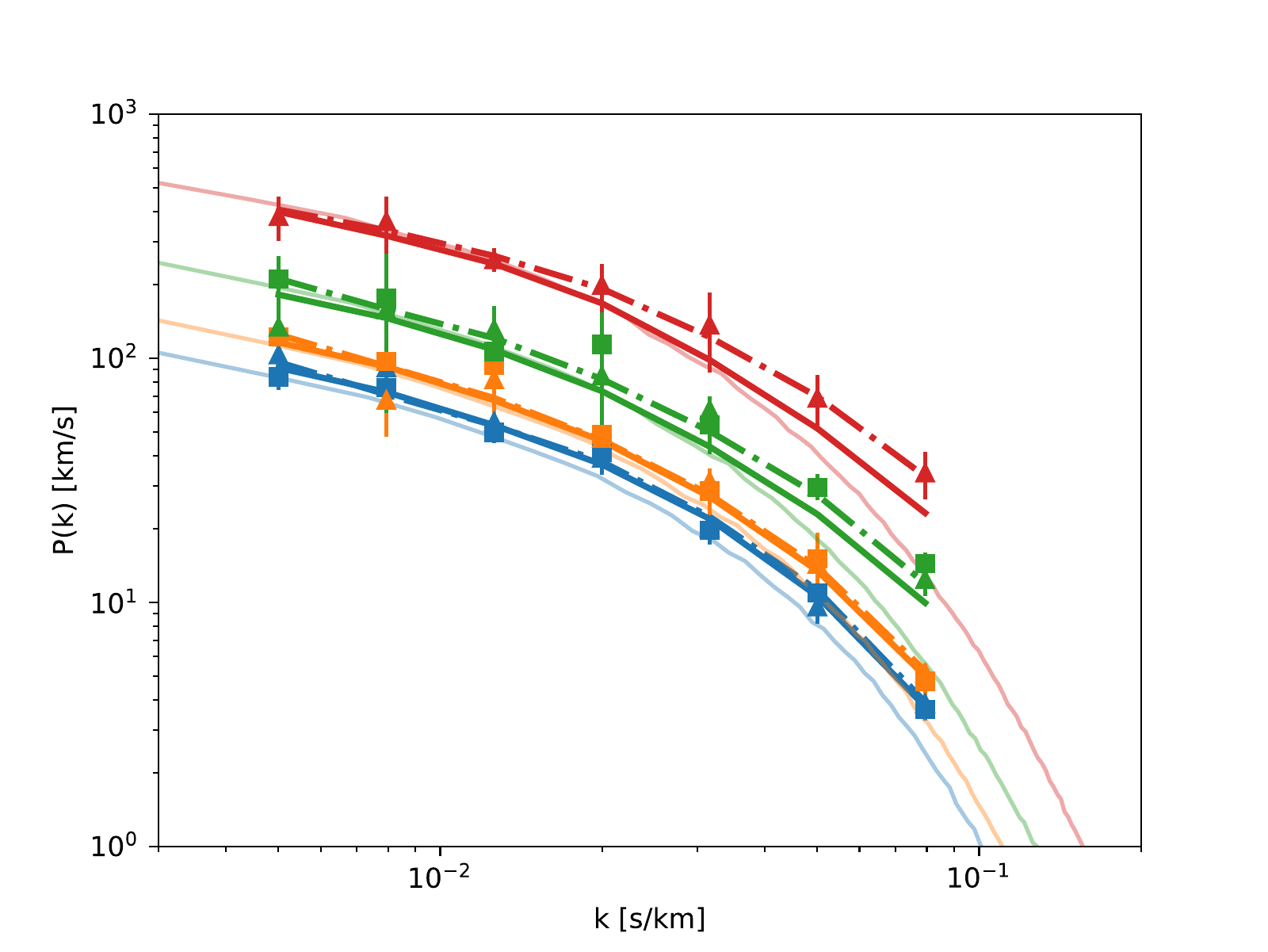}
	\caption{Model with $\sigma_{\mathrm{DM}-\mathrm{b}}=7.9 \cdot 10^{-27}$ and $f_{\mathrm{DM}-\mathrm{b}}=0.33$. The colors are $[z=4.2 - \mathrm{blue}, z=4.6 - \mathrm{orange}, z=5.0 - \mathrm{green}, z=5.4 - \mathrm{red}]$. The faint lines in the background are the additional simulation spectra themselves, which are then corrected using the astro/kriging grid in solid lines. The dashed lines are instead the predictions from the simulation grid.  The triangle correspond to HIRES data, while the squares correspond to MIKE data. \textbf{Left:} Bounded method. \textbf{Right:} Regularised method. \label{fig:compare_regbf}}
\end{figure}

In this case we ran additional simulations for two models of DM scattering with baryons with $n_b=0$, and the corresponding flux power spectra are shown in \cref{fig:compare_boubf,fig:compare_regbf} compared to the data and the simulated flux. The points were chosen to be strongly-interacting models (i.e. $\log(\sigma_{\mathrm{DM}-\mathrm{b}}) > -27$) and close to our best-fit regions, but otherwise randomly, to make sure the given preference is not caused by mispredictions of the flux spectra.

We observe very excellent agreement for the first point ($\sigma_{\mathrm{DM}-\mathrm{b}}= 6.2 \cdot 10^{-26}$ and $f_{\mathrm{DM}-\mathrm{b}}=0.23$ $\Rightarrow$ $\sigma_8 = 0.82$ and $n_\mathrm{eff} = -2.42$), while the second point ($\sigma_{\mathrm{DM}-\mathrm{b}}=7.9 \cdot 10^{-27}$ and $f_{\mathrm{DM}-\mathrm{b}}=0.33$ $\Rightarrow$ $\sigma_8 = 0.83$ and $n_\mathrm{eff} = -2.33$) still has reasonable agreement. In order to facilitate the comparison, we had to adjust the initial flux power spectra of the simulations with the same correction as for the original point from the original astro/cosmo grid (step 1 of the step-by-step description in \cref{ssec:interpolation}). As such, we primarily tested the validity of the suppression grid with this check, while the astro/cosmo corrections are assumed to be correct. 
\enlargethispage*{1\baselineskip}
\section{Additional details on the simulations}
In \cref{tab:abgdvalues} we list the parameters $\alpha,\beta,\gamma,\delta$ that were employed for each of the 200 simulations within the grid. Note that the \enquote{thermal} cases were run with given values of the mass, translating only roughly into a corresponding value of $\alpha$. Additionally, the simulations in \enquote{ABG} have mostly been chosen with a given value of $k_{1/2}$ (where $T(k_{1/2})=1/2$) in mind, resulting in the given values of $\alpha$. Finally, some of the \enquote{ABD} simulations have been performed in order to extend the existing grid in relevant sections, motivating the given $\alpha,\beta,\gamma,\delta$ values.
\begin{table}[h!]
	\centering
	\tiny
\resizebox{!}{10cm}{
	\begin{tabular}{c | c c c c || c | c c c c}
		Type & $\alpha$ [0.01Mpc/$h$] & $\beta$ & $\gamma$ & $\delta $ & Type & $\alpha$ [0.01Mpc/$h$]& $\beta$ & $\gamma$ & $\delta$ \\ \hline
ABG & 0.751 & 1.5 & -10.0 & 0 & Thermal & 2.270 & 2.24 & -4.464 & 0 \\
ABG & 0.479 & 1.5 & -10.0 & 0 & Thermal & 1.447 & 2.24 & -4.464 & 0 \\
ABG & 0.348 & 1.5 & -10.0 & 0 & Thermal & 1.052 & 2.24 & -4.464 & 0 \\
ABG & 1.206 & 1.5 & -5.0 & 0 & Thermal & 0.821 & 2.24 & -4.464 & 0 \\
ABG & 0.769 & 1.5 & -5.0 & 0 & Thermal & 0.671 & 2.24 & -4.464 & 0 \\
ABG & 0.559 & 1.5 & -5.0 & 0 & Thermal & 0.565 & 2.24 & -4.464 & 0 \\
ABG & 3.880 & 1.5 & -1.0 & 0 & Thermal & 0.487 & 2.24 & -4.464 & 0 \\
ABG & 2.474 & 1.5 & -1.0 & 0 & Thermal & 0.428 & 2.24 & -4.464 & 0 \\
ABG & 1.797 & 1.5 & -1.0 & 0 & ABD & 0.500 & 1.500 & -2.250 & 0.300 \\
ABG & 1.311 & 2.0 & -10.0 & 0 & ABD & 0.500 & 1.500 & -2.250 & 0.500 \\
ABG & 0.836 & 2.0 & -10.0 & 0 & ABD & 0.500 & 1.500 & -2.250 & 0.800 \\
ABG & 0.607 & 2.0 & -10.0 & 0 & ABD & 0.500 & 3.000 & -4.500 & 0.300 \\
ABG & 1.870 & 2.0 & -5.0 & 0 & ABD & 0.500 & 3.000 & -4.500 & 0.500 \\
ABG & 1.193 & 2.0 & -5.0 & 0 & ABD & 0.500 & 3.000 & -4.500 & 0.800 \\
ABG & 0.867 & 2.0 & -5.0 & 0 & ABD & 0.500 & 5.000 & -7.500 & 0.300 \\
ABG & 4.493 & 2.0 & -1.0 & 0 & ABD & 0.500 & 5.000 & -7.500 & 0.500 \\
ABG & 2.865 & 2.0 & -1.0 & 0 & ABD & 0.500 & 5.000 & -7.500 & 0.800 \\
ABG & 2.082 & 2.0 & -1.0 & 0 & ABD & 1.000 & 1.500 & -2.250 & 0.300 \\
ABG & 1.832 & 2.5 & -10.0 & 0 & ABD & 1.000 & 1.500 & -2.250 & 0.500 \\
ABG & 1.168 & 2.5 & -10.0 & 0 & ABD & 1.000 & 1.500 & -2.250 & 0.800 \\
ABG & 0.849 & 2.5 & -10.0 & 0 & ABD & 1.000 & 3.000 & -4.500 & 0.300 \\
ABG & 2.434 & 2.5 & -5.0 & 0 & ABD & 1.000 & 3.000 & -4.500 & 0.500 \\
ABG & 1.552 & 2.5 & -5.0 & 0 & ABD & 1.000 & 3.000 & -4.500 & 0.800 \\
ABG & 1.128 & 2.5 & -5.0 & 0 & ABD & 1.000 & 5.000 & -7.500 & 0.300 \\
ABG & 4.907 & 2.5 & -1.0 & 0 & ABD & 1.000 & 5.000 & -7.500 & 0.500 \\
ABG & 3.129 & 2.5 & -1.0 & 0 & ABD & 1.000 & 5.000 & -7.500 & 0.800 \\
ABG & 2.274 & 2.5 & -1.0 & 0 & ABD & 5.000 & 1.500 & -2.250 & 0.300 \\
ABG & 1.116 & 2.0 & -5.0 & 0 & ABD & 5.000 & 1.500 & -2.250 & 0.500 \\
ABG & 0.992 & 2.0 & -5.0 & 0 & ABD & 5.000 & 1.500 & -2.250 & 0.800 \\
ABG & 1.453 & 2.5 & -5.0 & 0 & ABD & 5.000 & 3.000 & -4.500 & 0.300 \\
ABG & 1.291 & 2.5 & -5.0 & 0 & ABD & 5.000 & 3.000 & -4.500 & 0.500 \\
ABG & 2.460 & 5.0 & -5.0 & 0 & ABD & 5.000 & 3.000 & -4.500 & 0.800 \\
ABG & 2.187 & 5.0 & -5.0 & 0 & ABD & 5.000 & 5.000 & -7.500 & 0.300 \\
ABG & 3.202 & 10.0 & -5.0 & 0 & ABD & 5.000 & 5.000 & -7.500 & 0.500 \\
ABG & 2.846 & 10.0 & -5.0 & 0 & ABD & 5.000 & 5.000 & -7.500 & 0.800 \\
ABG & 9.527 & 2.5 & -0.3 & 0 & ABD & 10.000 & 1.500 & -2.250 & 0.300 \\
ABG & 16.873 & 2.5 & -0.15 & 0 & ABD & 10.000 & 1.500 & -2.250 & 0.500 \\
ABG & 6.074 & 2.5 & -0.3 & 0 & ABD & 10.000 & 1.500 & -2.250 & 0.800 \\
ABG & 10.759 & 2.5 & -0.15 & 0 & ABD & 10.000 & 3.000 & -4.500 & 0.300 \\
ABG & 4.414 & 2.5 & -0.3 & 0 & ABD & 10.000 & 3.000 & -4.500 & 0.500 \\
ABG & 7.817 & 2.5 & -0.15 & 0 & ABD & 10.000 & 3.000 & -4.500 & 0.800 \\
ABG & 5.685 & 2.5 & -0.3 & 0 & ABD & 10.000 & 5.000 & -7.500 & 0.300 \\
ABG & 10.070 & 2.5 & -0.15 & 0 & ABD & 10.000 & 5.000 & -7.500 & 0.500 \\
ABG & 5.054 & 2.5 & -0.3 & 0 & ABD & 10.000 & 5.000 & -7.500 & 0.800 \\
ABG & 8.951 & 2.5 & -0.15 & 0 & ABD & 100.000 & 1.000 & -1.500 & 0.300 \\
ABG & 8.155 & 5.0 & -0.3 & 0 & ABD & 100.000 & 1.000 & -1.500 & 0.500 \\
ABG & 10.854 & 5.0 & -0.15 & 0 & ABD & 100.000 & 1.000 & -1.500 & 0.700 \\
ABG & 5.200 & 5.0 & -0.3 & 0 & ABD & 100.000 & 1.500 & -2.250 & 0.300 \\
ABG & 6.921 & 5.0 & -0.15 & 0 & ABD & 100.000 & 1.500 & -2.250 & 0.500 \\
ABG & 3.779 & 5.0 & -0.3 & 0 & ABD & 100.000 & 1.500 & -2.250 & 0.700 \\
ABG & 5.029 & 5.0 & -0.15 & 0 & ABD & 500.000 & 1.000 & -1.500 & 0.300 \\
ABG & 4.867 & 5.0 & -0.3 & 0 & ABD & 500.000 & 1.000 & -1.500 & 0.500 \\
ABG & 6.477 & 5.0 & -0.15 & 0 & ABD & 500.000 & 1.000 & -1.500 & 0.700 \\
ABG & 4.326 & 5.0 & -0.3 & 0 & ABD & 500.000 & 1.500 & -2.250 & 0.300 \\
ABG & 5.758 & 5.0 & -0.15 & 0 & ABD & 500.000 & 1.500 & -2.250 & 0.500 \\
ABG & 2.343 & 2.0 & -6.0 & 0 & ABD & 500.000 & 1.500 & -2.250 & 0.700 \\
ABG & 0.915 & 2.0 & -6.0 & 0 & ABD & 0.050 & 0.500 & -0.750 & 0.000 \\
ABG & 0.616 & 2.0 & -6.0 & 0 & ABD & 0.050 & 0.750 & -1.125 & 0.000 \\
ABG & 2.891 & 2.0 & -4.0 & 0 & ABD & 0.050 & 1.000 & -1.500 & 0.000 \\
ABG & 1.129 & 2.0 & -4.0 & 0 & ABD & 0.100 & 0.500 & -0.750 & 0.000 \\
ABG & 0.760 & 2.0 & -4.0 & 0 & ABD & 0.100 & 0.750 & -1.125 & 0.000 \\
ABG & 4.180 & 2.0 & -2.0 & 0 & ABD & 0.100 & 1.000 & -1.500 & 0.000 \\
ABG & 1.632 & 2.0 & -2.0 & 0 & ABD & 0.500 & 0.500 & -0.750 & 0.000 \\
ABG & 1.098 & 2.0 & -2.0 & 0 & ABD & 0.500 & 0.750 & -1.125 & 0.000 \\
ABG & 4.745 & 4.0 & -6.0 & 0 & ABD & 0.500 & 1.000 & -1.500 & 0.000 \\
ABG & 1.853 & 4.0 & -6.0 & 0 & ABD & 1.000 & 0.500 & -0.750 & 0.000 \\
ABG & 1.247 & 4.0 & -6.0 & 0 & ABD & 1.000 & 0.750 & -1.125 & 0.000 \\
ABG & 5.270 & 4.0 & -4.0 & 0 & ABD & 1.000 & 1.000 & -1.500 & 0.000 \\
ABG & 2.058 & 4.0 & -4.0 & 0 & ABD & 1.225 & 1.680 & -2.520 & 0.857 \\
ABG & 1.385 & 4.0 & -4.0 & 0 & ABD & 6.935 & 2.047 & -3.070 & 0.884 \\
ABG & 6.337 & 4.0 & -2.0 & 0 & ABD & 29.813 & 2.021 & -3.031 & 0.884 \\
ABG & 2.474 & 4.0 & -2.0 & 0 & ABD & 117.595 & 1.853 & -2.780 & 0.874 \\
ABG & 1.665 & 4.0 & -2.0 & 0 & ABD & 1.244 & 1.667 & -2.501 & 0.754 \\
ABG & 6.003 & 6.0 & -6.0 & 0 & ABD & 6.904 & 2.020 & -3.030 & 0.798 \\
ABG & 2.344 & 6.0 & -6.0 & 0 & ABD & 30.708 & 1.994 & -2.991 & 0.799 \\
ABG & 1.578 & 6.0 & -6.0 & 0 & ABD & 119.849 & 1.844 & -2.766 & 0.784 \\
ABG & 6.439 & 6.0 & -4.0 & 0 & ABD & 1.242 & 1.637 & -2.455 & 0.580 \\
ABG & 2.514 & 6.0 & -4.0 & 0 & ABD & 7.032 & 1.992 & -2.988 & 0.657 \\
ABG & 1.692 & 6.0 & -4.0 & 0 & ABD & 30.842 & 1.977 & -2.965 & 0.659 \\
ABG & 7.281 & 6.0 & -2.0 & 0 & ABD & 122.085 & 1.835 & -2.752 & 0.637 \\
ABG & 2.843 & 6.0 & -2.0 & 0 & ABD & 1.330 & 1.634 & -2.451 & 0.332 \\
ABG & 1.913 & 6.0 & -2.0 & 0 & ABD & 7.298 & 1.945 & -2.918 & 0.440 \\
ABG & 1.971 & 3.0 & -7.5 & 0 & ABD & 32.732 & 1.930 & -2.894 & 0.444 \\
ABG & 1.026 & 3.0 & -7.5 & 0 & ABD & 129.793 & 1.829 & -2.743 & 0.419 \\
ABG & 0.913 & 3.0 & -7.5 & 0 & ABD & 1.556 & 1.650 & -2.475 & 0.057 \\
ABG & 2.888 & 3.0 & -2.5 & 0 & ABD & 8.109 & 1.886 & -2.830 & 0.162 \\
ABG & 1.504 & 3.0 & -2.5 & 0 & ABD & 35.322 & 1.873 & -2.809 & 0.170 \\
ABG & 1.338 & 3.0 & -2.5 & 0 & ABD & 146.441 & 1.815 & -2.723 & 0.155 \\
ABG & 4.063 & 3.0 & -1.0 & 0 & ABD & 2.402 & 1.912 & -2.869 & 0.000 \\
ABG & 2.116 & 3.0 & -1.0 & 0 & ABD & 850.000 & 2.000 & -3.000 & 0.825 \\
ABG & 1.882 & 3.0 & -1.0 & 0 & ABD & 1000.000 & 1.500 & -2.250 & 0.925 \\
ABG & 2.961 & 5.0 & -7.5 & 0 & ABG & 3.525 & 7.0 & -7.5 & 0 \\
ABG & 1.542 & 5.0 & -7.5 & 0 & ABG & 1.836 & 7.0 & -7.5 & 0 \\
ABG & 1.372 & 5.0 & -7.5 & 0 & ABG & 1.633 & 7.0 & -7.5 & 0 \\
ABG & 3.723 & 5.0 & -2.5 & 0 & ABG & 4.151 & 7.0 & -2.5 & 0 \\
ABG & 1.939 & 5.0 & -2.5 & 0 & ABG & 2.162 & 7.0 & -2.5 & 0 \\
ABG & 1.725 & 5.0 & -2.5 & 0 & ABG & 1.923 & 7.0 & -2.5 & 0 \\
ABG & 4.570 & 5.0 & -1.0 & 0 & ABG & 4.806 & 7.0 & -1.0 & 0 \\
ABG & 2.380 & 5.0 & -1.0 & 0 & ABG & 2.503 & 7.0 & -1.0 & 0 \\
ABG & 2.117 & 5.0 & -1.0 & 0 & ABG & 2.226 & 7.0 & -1.0 & 0 \\
	\end{tabular}
}
	\caption{\label{tab:abgdvalues} Values of the $\alpha$, $\beta$, $\gamma$, and $\delta$ parameters for all 200 suppression simulations used within this work.}
\end{table}